\documentclass[journal,twocolumn]{IEEEtran}

\usepackage[final]{graphicx}

\usepackage{amsmath,epsfig,amssymb,verbatim,amsopn,cite,subfigure,multirow,amsthm}
\usepackage{balance}
\usepackage[usenames,dvipsnames]{color}
\usepackage[all]{xy}  
\usepackage{url}
\usepackage{amsfonts}
\usepackage{amssymb}
\usepackage{epsfig}
\usepackage{bm}
\usepackage{epsfig}
\usepackage{amsmath,amssymb, amstext}
\usepackage{graphicx}
\usepackage{epstopdf}
\usepackage{algorithm}
\usepackage{algorithmicx}
\usepackage{tabularx,booktabs}
\usepackage{url}
\usepackage{multirow}
\usepackage{array}
\usepackage{makecell}
\usepackage{balance}
\usepackage{algorithm}
\usepackage{algpseudocode}
\usepackage{bbm}
\usepackage{algcompatible}
\usepackage{booktabs,arydshln}
\usepackage{subfig}
\usepackage{caption}
\usepackage{algorithm}
\usepackage{algpseudocode}
\usepackage{algcompatible}

\newtheorem{lemma}{Lemma}

\usepackage{accents}
\newlength{\dhatheight}

\IEEEaftertitletext{\vspace{-2.5\baselineskip}}

\begin{document}
	
	\title{Packet Loss Analysis Under Doppler Effects and Reinforcement-Learning-Based Resource Allocation for Direct-to-Satellite LoRaWAN 
	}
	
	\author{Alireza Maleki, Ebrahim Bedeer, and Robert Barton
		\thanks{Alireza Maleki and Ebrahim Bedeer are with the Department of Electrical and Computer Engineering, University of Saskatchewan, Saskatoon, Canada S7N5A9. Emails: \{alireza.maleki and e.bedeer\}@usask.ca.}
		\thanks{R. Barton is with Cisco Systems Inc. Email: robbarto@cisco.com.}
		\thanks{This work was supported by the NSERC/Cisco Industrial Research Chair program.}
	}
	\maketitle
	
	\begin{abstract}
	Long range wide area network (LoRaWAN) has a long communication range and low power consumption, making it a potential technology for direct-to-satellite (DtS) Internet of Things (IoT) connectivity. However, packet collisions and Doppler impairments brought on by satellite motion can greatly increase the packet loss ratio (PLR) in large-scale DtS-IoT networks. In this paper, we study resource allocation in DtS-IoT LoRaWAN by first developing an analytical PLR model that simultaneously takes Doppler impairments and packet collisions in the temporal and frequency domains into account. In order to maximize resource allocation, the generated PLR is then used as the reward function for two reinforcement learning (RL) schemes: deep Q-network (DQN) and proximal policy optimization (PPO). Simulation results validate the analytical PLR model against Monte Carlo simulations and demonstrate significant improvements over random resource allocation. At a target PLR of $3\times10^{-2}$, while random allocation supports only $100$ end devices (EDs), PPO and DQN support approximately $650$ and $690$ EDs, respectively, representing nearly a seven-fold increase. Furthermore, for $800$ EDs, PPO and DQN increase the total number of transmitted packets by $106.7\%$ and $88.0\%$, respectively, compared with random allocation.
	\end{abstract}
	
	\section{Introduction}
	\label{intro}
	\IEEEPARstart{I}{n} recent years, the coverage limitations of terrestrial low-power wide-area networking (LPWAN) systems have been revealed by the growing deployment of IoT end devices (EDs), especially in remote and large-scale situations such as offshore wind farms \cite{Ullah_2021,Maleki_D2D}. Consequently, direct-to-satellite IoT (DtS-IoT), particularly through low Earth orbit (LEO) satellites, has emerged as a feasible means of providing wide-area coverage and connectivity to many EDs, including those in remote and underserved locations \cite{ref_1_PLR,TVT_Intro1}. Several LPWAN technologies, including long range wide area network (LoRaWAN) and narrow-band IoT (NB-IoT), have been investigated for low-power DtS-IoT connectivity \cite{Doroshkin_2019,Ouvry_2018,Deng_2017}. Commercial initiatives such as Wyld\footnote{https://www.wyldnetworks.com/} and Lacuna Space\footnote{https://lacuna.space/} are also developing LEO satellite constellations incorporating LoRaWAN gateways (GWs) for global IoT connectivity \cite{Alvarez_2022,Afhamisis_2022}. Furthermore, integrating LPWAN technologies with nano-satellites can provide a cost-effective and delay-tolerant alternative to conventional satellite systems for IoT services \cite{Almonacid_2017,Akyildiz_2019}.
	
	Among LPWAN technologies exploited in DtS-IoT, LoRaWAN is one of the most popular ones that allows low-power and long-range connectivity. Because of its cheap deployment cost, energy efficiency, and capacity to serve a high number of EDs over wide coverage areas, LoRaWAN, which is based on the LoRa modulation \cite{Maleki_chirp}, is seen as a promising technology for DtS-IoT networks. The performance of LoRaWAN over satellite links has thus been the subject of extensive research, with previous studies examining elements like link budget, interference, and communication reliability, proving the viability of direct communications between LoRaWAN devices and LEO satellites \cite{Doroshkin_2019,Ullah_2021,Ullah_2022,Asad_2024}.
	
	However, there are a number of technical difficulties with integrating LoRaWAN with LEO satellite networks, most notably the Doppler effect brought on by the high orbital velocity of LEO satellites. Because LoRa signals are sensitive to frequency changes, the ensuing time-varying Doppler shift and Doppler rate can have a substantial impact on communication dependability \cite{Asad_2024}. As a result, Doppler effects in LoRa communications have been the subject of numerous investigations. The authors of \cite{Asad_2024} examined how Doppler shift and rate affected DtS-IoT LoRaWAN's packet loss performance at various bandwidths, carrier frequencies, payload sizes, and orbital altitudes. An SDR-based experimental platform was used in \cite{LoRa_Dopp_1} to assess LoRa performance under various acceleration and speed situations, offering insights into its suitability for Doppler-sensitive scenarios. Additionally, \cite{LoRa_Dopp_2} examined the performance and parameter trade-offs of four Doppler estimation and compensation frameworks for DtS-IoT LoRaWAN in comparison to an ideal Doppler-free environment.
	
	Allocating resources for a large number of EDs across the vast coverage area of a satellite-mounted GW is another significant difficulty in DtS-IoT LoRaWAN. The traditional ALOHA-based random access of LoRaWAN is ineffective since transmissions from several EDs may overlap in both time and frequency, leading to serious packet collisions \cite{SX1272}. Semtech developed long range frequency hopping spread spectrum (LR-FHSS), which uses narrowband frequency-hopping transmissions and packet fragmentation to lessen the likelihood of collisions \cite{9,Ullah_2021,Maleki_2023,Knop_2024}. However, in contrast to LoRa, LR-FHSS transmits packet fragments using Gaussian minimum shift keying (GMSK) \cite{Maleki_D2D}. By taking into consideration satellite footprint geometry and orbital motion, \cite{Testi_2024} analytically analyzed packet collisions in DtS-IoT and showed that LR-FHSS can sustain significantly more simultaneous broadcasts than traditional LoRaWAN.
		
	Machine learning (ML) has emerged as a promising approach for solving complex resource allocation problems in wireless communication systems, including power control, spectrum allocation, and user scheduling \cite{Surv_ML_RA}. As a result, ML has been used in a number of experiments to allocate resources in LoRaWAN. A TinyML-based channel allocation technique was presented in \cite{TinyML} to allow autonomous channel selection at EDs, increasing the packet success percentage and decreasing inference time and model size. In \cite{CORA}, a hybrid deep learning and convex optimization framework was developed for joint channel, spreading factor, and transmit power allocation, achieving near-optimal resource allocation and improved energy efficiency. Moreover, \cite{AI_Tiny} proposed a hybrid TinyML and deep learning framework for joint spreading factor and transmit power allocation, improving packet success ratio and energy efficiency for both static and mobile IoT applications.
	According the above-mentioned discussions, the following shortcomings can be listed regarding the existing literature:
	\begin{itemize}
		\item The majority of the current research in the performance analysis of DtS-IoT LoRaWAN has relied on simulation-based assessments or experimental measurements.
		\item To the best of the authors' knowledge, existing analytical studies on DtS-IoT LoRaWAN have not provided a comprehensive framework for evaluating the packet loss ratio (PLR) by jointly accounting for Doppler impairments and random resource allocation.
		\item As a proven and promising solution in resource allocation, ML approaches have not been investigated for DtS-IoT LoRaWAN while capturing both Doppler and collision effects. 
	\end{itemize}

	To address these challenges, we develop a comprehensive DtS-IoT LoRaWAN system model that captures the effects of random resource allocation, Doppler shift, and Doppler rate on the PLR of EDs served by a satellite-mounted LoRa GW. Unlike existing studies that primarily rely on experiments or simulations, we analytically characterize the PLR by jointly modeling random access and Doppler impairments due to satellite motion and validate the derived expression through Monte Carlo simulations. Since joint resource allocation among all EDs while accounting for time-varying Doppler effects results in a computationally intractable optimization problem, we employ the derived PLR as the reward function for reinforcement learning (RL)-based resource allocation. The trained RL agents learn efficient resource allocation policies to improve the overall network performance. Accordingly, the main contributions of this work are summarized as follows:
	\begin{enumerate}
		\item We present a detailed analytical procedure to obtain a mathematical expression for the PLR of a typical DtS-IoT LoRaWAN scheme that captures the packet loss due to collision in time and frequency domains, Doppler shift, and Doppler rate.
		\item Building on these derivations, we reframe the resource allocation problem as an RL task using proximal policy optimization (PPO) and deep Q-network (DQN) algorithms where each ED's $\rm SF$, $\rm CR$, $\rm BW$, and first packet start time are assigned by an agent. Importantly, the mathematical PLR expression obtained in the first section serves as both a performance metric and the design foundation for the RL formulation itself. The construction of the state features and the per-packet reward function is directly guided by the analytical dependence of the PLR on the transmission parameters, which reveals how each resource affects Doppler shift, Doppler rate, and collision losses. This ensures that the learning objective is directly linked to the underlying physical loss mechanisms by using the derived PLR as the reward signal that trains the agent.
		\item Using the PPO and DQN algorithms, we implement the presented RL framework and assess it using computer simulations against a baseline of random resource allocation. The findings demonstrate that both learned policies significantly lower the PLR for all network sizes, and for a target PLR of $3\times10^{-2}$, enable serving roughly $6.5\times$ and $6.9\times$ as many EDs as the random allocation for PPO and DQN, respectively. Concurrently, exploiting PPO and DQN resource allocation schemes increases the number of transmitted packets in the network compared to the random scheme, e.g., $106.7\%$ and  $88.0\%$ more packets for $800$ EDs, respectively.
	\end{enumerate}  
	
	The rest of the paper is organized as follows. In Section~\ref{sysmod}, the system model is presented. Section~\ref{plr} provides the analytical derivation of the PLR performance. In Section~\ref{sim}, the simulation results are given. Finally, in Section~\ref{conc}, the paper is concluded. 
	
	\section{System Model}
	\label{sysmod}
\begin{figure}[t!]
		\centering
		\includegraphics[width=0.6\linewidth]{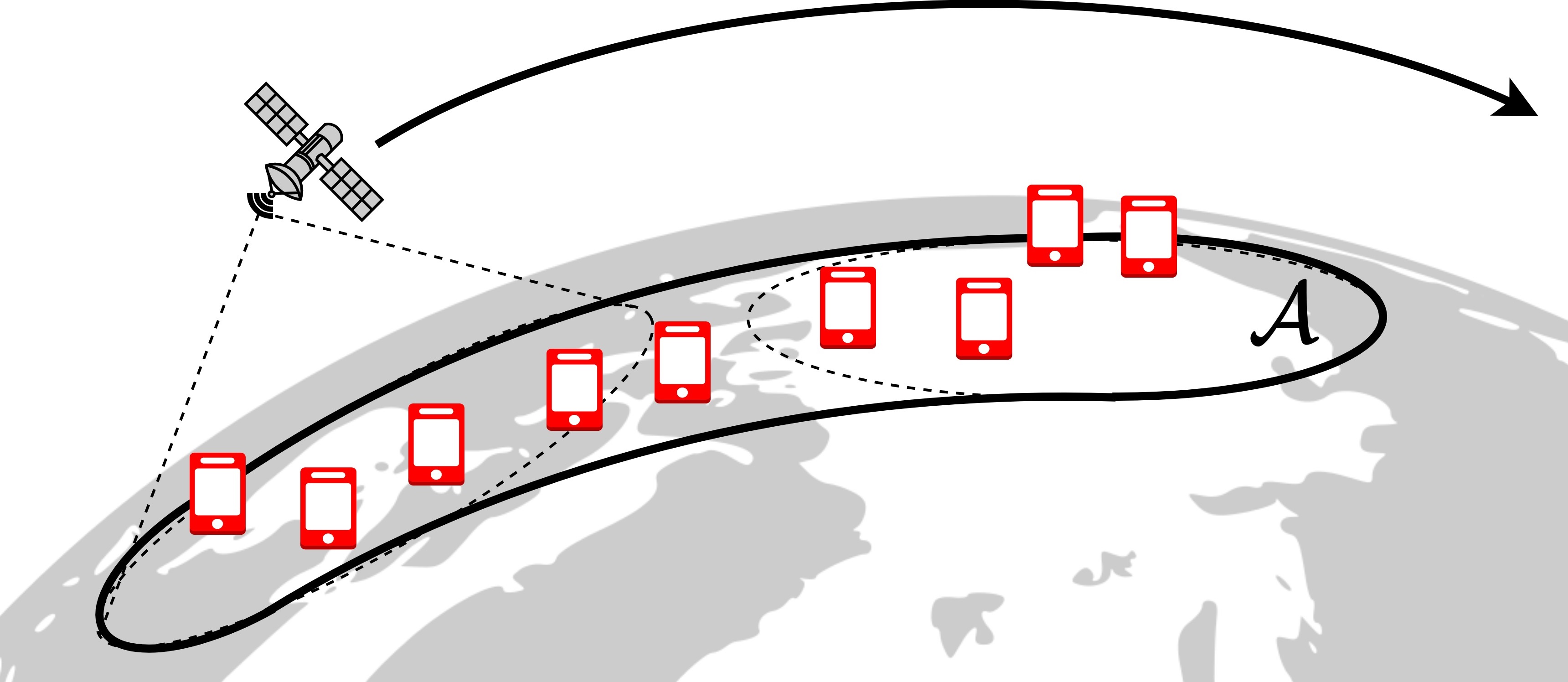}
		\caption{LoRa DtS-IoT network under consideration.}
		\label{DtS_Net}
	\end{figure}

	We consider a network of $N$ terrestrial IoT end-devices (EDs) transmitting their packets to a LoRaWAN GW installed on a LEO satellite. We assume that these EDs are distributed randomly via a uniform distribution within the coverage area $\mathcal{A}$ as illustrated schematically in Fig. \ref{DtS_Net}. In the following, we present the LoRaWAN and satellite parameters associated with the DtS-IoT network.
	
	\subsection{LoRaWAN Parameters}
We assume that EDs utilize chirp spread spectrum (CSS) modulation for packet transmission \cite{Maleki_chirp}. Accordingly, based on LoRaWAN specifications, before the start of a packet transmission, each ED selects three different transmission parameters as follows according to \cite{Things_LoRa_Par} (for now, we assume that these selections are random):
	\begin{enumerate}
		\item Spreading factor (${\rm SF}\in\{7,8,9,10\}$),
		\item Coding rate (${\rm CR}\in\{4/5,4/6,4/7,4/8\}$),
		\item Channel (${\rm CH} \in\{1,2,\dots,64\}$).  
	\end{enumerate}
	Moreover, we assume that the transmission power $P_{\rm t}$, bandwidth $\rm BW$, and the payload size ${\rm PL}$ in bytes are the same for each ED. Note that these assumptions simplify the derivation, are made solely for analytical convenience, and would not affect the generality of the presented derivation. Based on these parameters, we can calculate the time-on-air (ToA) of a single packet transmitted by $n$th ED ($n=1,2,\dots,N$) as:	 
	\begin{IEEEeqnarray}{rCl}
		\label{ToA}
		{\rm ToA}_n = \frac{{\rm PL}\times 8 \times T_s}{{\rm SF}\times {\rm CR}}, 
	\end{IEEEeqnarray}
	where $T_s=2^{\rm SF}/{\rm BW}$ is the CSS symbol duration \cite{Maleki_chirp}. 
	\subsection{Satellite Parameters}
\begin{figure}[t!]
	\centering
	\includegraphics[width=0.6\linewidth]{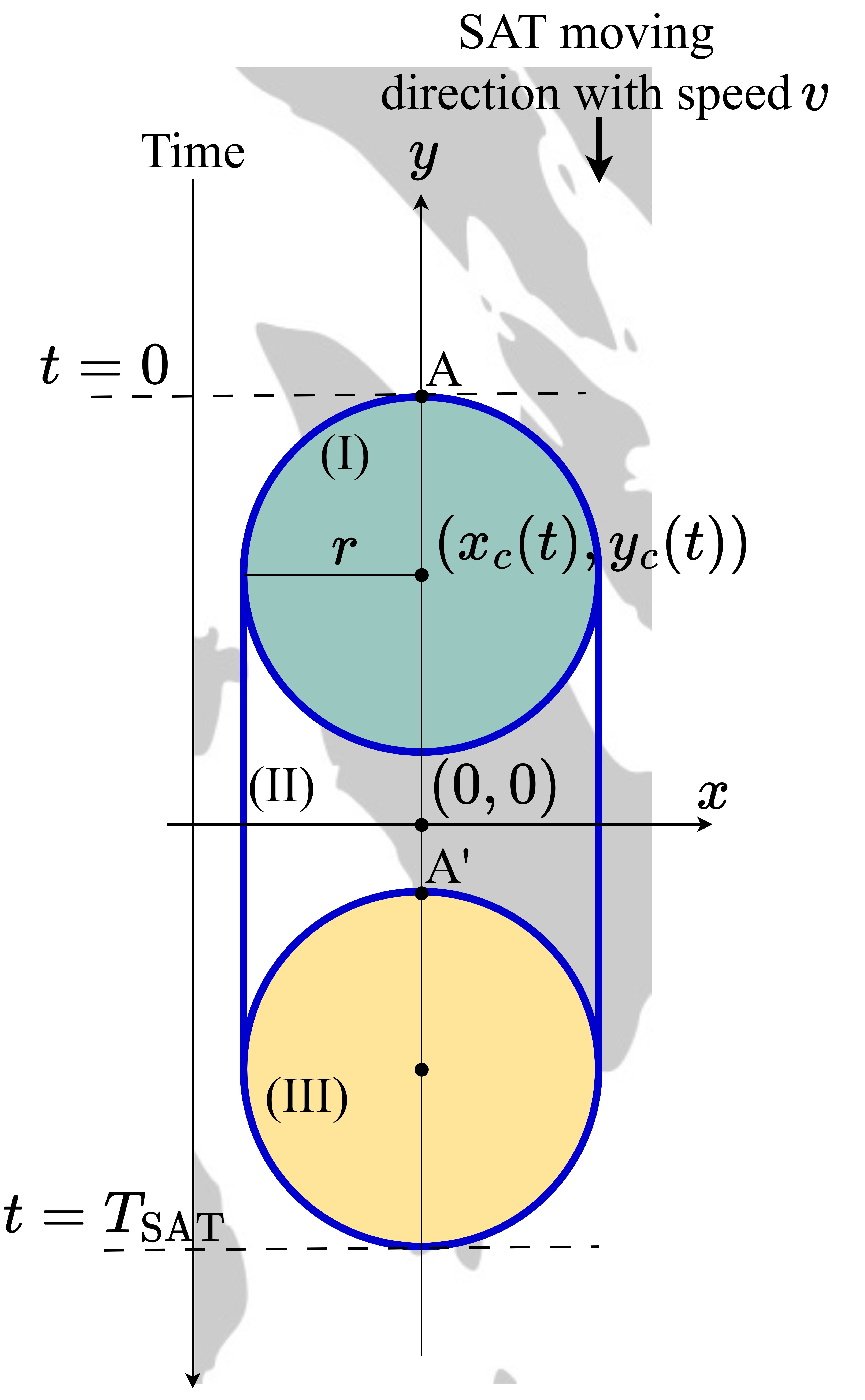}
	\caption{Illustration of the top view of the covered area by the satellite.}
	\label{TopView}
\end{figure}

	To better understand the geometry of the satellite, we illustrate the top-view of the coverage area $\mathcal{A}$ in Fig. \ref{TopView}. As can be seen, we assume that the satellite starts moving at $t=0$ and reaches the end of the coverage area at $t=T_{\rm SAT}$. Also, $r$ is the satellite footprint radius which can be calculated as \cite{Asad_2024}:
	\begin{IEEEeqnarray}{rCl}
		\label{r_SAT}
		r = R\sin^{-1}\left[\frac{\zeta(E_{\rm min})\cos(E_{\rm min})}{R+H}\right], 
	\end{IEEEeqnarray}
	where $E_{\rm min}$ is the minimum elevation angle of the satellite ($1$ degree in the most simple case), $R$ is the radius of Earth, $H$ denotes the satellite orbit height, and
	\begin{IEEEeqnarray}{rCl}
		\label{zetaE}
		\zeta(E) = R\left[\sqrt{\left(\frac{H+R}{R}\right)^2-\cos^2(E)}-\sin(E)\right].
	\end{IEEEeqnarray}
Moreover, note that the center of the satellite footprint coordinates can be represented as a function of time and based on our geometry as $(x_c(t),y_c(t))=(0,y_c(0)-vt)$ where $v$ is the velocity of the satellite in the circular orbit and can be computed by $v=\sqrt{gH/(1+H/R)}$ where $g$ is the gravitational acceleration of Earth \cite{Asad_2024}.

Considering that the $N$ EDs are randomly and uniformly distributed in the area shown in Fig. \ref{TopView}, each ED can transmit its packet(s) only when it is visible by the satellite. Based on this, it is worth noting that the number of packets transmitted by an ED is inherently location-dependent. Specifically, an ED located on the trajectory of the satellite footprint center in Area II experiences the maximum visibility time and, consequently, can transmit the largest possible number of packets. In contrast, as the ED is located toward the boundary of the coverage area, the visibility time decreases. For some edge-located EDs, the available visibility time may even be insufficient for transmitting a single packet. Therefore, the number of packets that an ED can transmit is a direct function of its location within the satellite coverage area. Depending on the location of the ED (whether it is in Area (I), (II), or (III) according to Fig. \ref{TopView}), the following parameters can be calculated for an arbitrary ED, e.g., ${\rm ED}_n$, with coordinates $(x_n,y_n)$:
\subsubsection{Visibility distance $d_n$} This is the distance traveled by the satellite while it remains visible to ${\rm ED}_n$. This corresponds to the distance traveled by the satellite from the moment its footprint first reaches the location of ${\rm ED}_n$ until it leaves that location. This distance can be calculated as follows:

\begin{IEEEeqnarray}{rCl}
	\label{dn}
	d_{n}=\begin{cases}
		y_c(0)-y_n+\sqrt{r^2-x^2_n}, & \text{if $(x_n,y_n)\in$ Area (I)},\\
		2\sqrt{r^2-x^2_n}, & \text{if $(x_n,y_n)\in$ Area (II)},\\
		y_c(0)+y_n+\sqrt{r^2-x^2_n}, & \text{if $(x_n,y_n)\in$ Area (III)}.
	\end{cases}\nonumber\\
\end{IEEEeqnarray}

\subsubsection{Transmission time parameters}
Regarding the time-related transmission parameters for ${\rm ED}_n$, the following parameters are defined:
\begin{itemize}
	\item Starting time of satellite visibility, i.e., the time instant that the satellite footprint reaches ${\rm ED}_n$'s location:
\begin{IEEEeqnarray}{rCl}
	\label{tsn}
	t_{{\rm s},n}=\begin{cases}
		0, & \text{if $(x_n,y_n)\in$ Area (I)},\\
		\frac{y_c(0)-y_n-\sqrt{r^2-x_n^2}}{v}, & \text{o.w.}.
	\end{cases}.
\end{IEEEeqnarray}
\item Satellite visibility duration:
\begin{IEEEeqnarray}{rCl}
	\label{Tvn}
	t_{{\rm v},n} = \frac{d_n}{v}.
\end{IEEEeqnarray}
\item Ending time of satellite visibility, i.e., the time instant that the satellite footprint leaves ${\rm ED}_n$'s location:
\begin{IEEEeqnarray}{rCl}
	\label{ten}
	t_{{\rm e},n} = t_{{\rm s},n}+t_{{\rm v},n}.
\end{IEEEeqnarray}
\end{itemize}	
It should be noted that for ${\rm ED}_n$, due to LoRaWAN specifications \cite{9}, a duty cycle constraint of $1\%$ must be met. Hence, the available time for transmission is considered as $0.01\times t_{{\rm v},n}$ and ${\rm ED}_n$ can transmit up to $N_{\rm p}(t_{{\rm v},n},{\rm ToA}_n)=\lfloor 0.01 t_{{\rm v},n}/{\rm ToA}_n\rfloor$ LoRa packets while visible to the satellite.

\subsubsection{Doppler shift}
\begin{figure}[t!]
	\centering
	\includegraphics[width=0.5\linewidth]{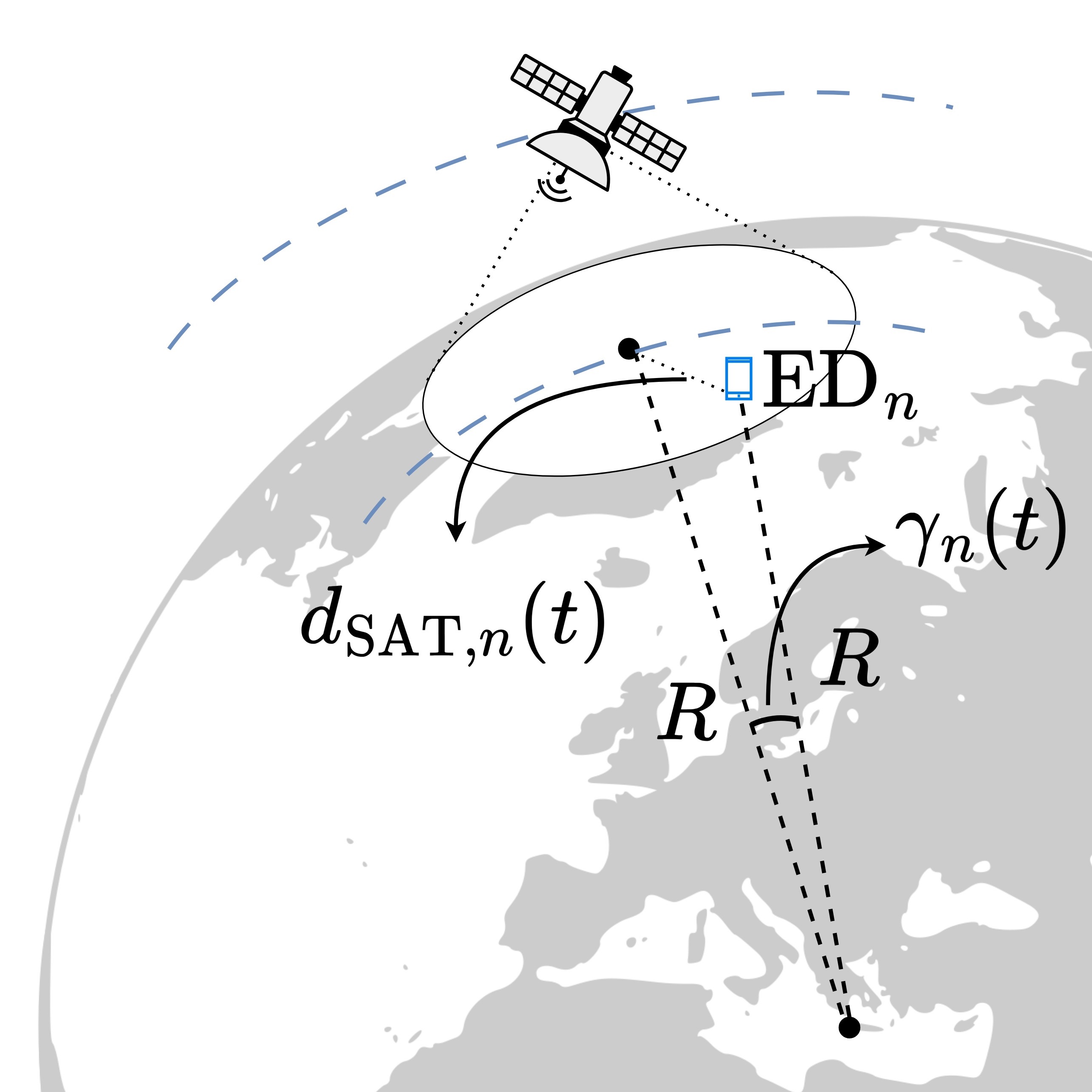}
	\caption{Illustration of the central angle $\gamma_n(t)$ for ${\rm ED}_n$.}
	\label{Dopp_Geo}
\end{figure}

Consider the distance between ${\rm ED}_n$'s location and the center of satellite footprint as:
\begin{IEEEeqnarray}{rCl}
	\label{dSATv1}
	d_{{\rm SAT},n}(t)&=&\sqrt{[x_n-x_c(t)]^2+[y_n-y_c(t)]^2}\nonumber\\
	&=&\sqrt{x_n^2+(y_n-y_c(0)+vt)^2},
\end{IEEEeqnarray}
which obviously is a function of time. Based on this, we define the central angle between ${\rm ED}_n$'s location, satellite footprint center, and the center of Earth, $\gamma_n(t)$ as shown in Fig. \ref{Dopp_Geo}. It is straightforward to show that we have:
\begin{IEEEeqnarray}{rCl}
	\label{gamma}
	\gamma_{n}(t) = 2\sin^{-1}\left[\frac{d_{{\rm SAT},n}(t)}{2R}\right].
\end{IEEEeqnarray}

Assuming that ${\rm ED}_n$ exploits frequency $f_0$ for its packet transmission, the Doppler shift magnitude experienced by the GW regarding the transmission of ${\rm ED}_n$ can be formulated as \cite{Tanash_2025}
\begin{IEEEeqnarray}{rCl}
	\label{dopp}
	\xi_n(t) = \rho\sqrt{\frac{\cos^2(\gamma_{{\rm min},n})-\cos^2\left[\gamma_n(t)\right]}{1+K^2-2K\cos\left[\gamma_n(t)\right]}},
\end{IEEEeqnarray}
where $\rho=f_0 R \omega /c$ with $\omega=v/(R+H)$ as the satellite angular velocity in Earth-centered fixed (ECF) frame \cite{Tanash_2025}, and $c$ as the speed of light. Moreover, $K=R/(R+H)$ and $\gamma_{{\rm min},n}$ is the minimum of the central angle $\gamma_n(t)$ within the duration $[t_{{\rm s},n},t_{{\rm e},n}]$. From (\ref{dopp}), it can be observed that the Doppler shift $\xi_n(t)$ depends on both the instantaneous central angle $\gamma_n(t)$ and the minimum central angle $\gamma_{{\rm min},n}$, which represents the closest approach of the satellite footprint center to the ED. At the closest approach, i.e., when $\gamma_n(t)=\gamma_{{\rm min},n}$, the numerator of (\ref{dopp}) becomes zero; hence, the Doppler shift is zero at this instant. Moreover, EDs located close to the trajectory of the satellite footprint center experience smaller $\gamma_{{\rm min},n}$ as it is straightforward to show that $\gamma_{{\rm min},n}=2\sin^{-1}[|x_n|/(2R)]$. Therefore, the term $\cos^2(\gamma_{{\rm min},n})-\cos^2[\gamma_n(t)]$ can become larger for these EDs, resulting in a larger Doppler shift range during the visibility interval. In contrast, EDs located close to the coverage boundary have larger $\gamma_{{\rm min},n}$, which limits the variation of $\gamma_n(t)$ during visibility and leads to smaller Doppler shift values. Therefore, although edge EDs experience shorter visibility times, the EDs located closer to the satellite ground-track are more vulnerable to Doppler-shift-induced packet loss, especially near the beginning and end of their visibility intervals.

Based on these parameters, in the following section, we derive the PLR for our system model.

\section{Packet Loss Ratio Derivation}
\label{plr}
To derive the total PLR for the presented network, we divide the problem into three separate packet loss events as packet loss due to Doppler shift, i.e., $\mathcal{E}_{\rm DS}$; packet loss due to Doppler rate, i.e., $\mathcal{E}_{\rm DR}$; and packet loss due to collision, i.e., $\mathcal{E}_{\rm Col}$.
\subsection{Packet loss due to Doppler shift}
As illustrated in \cite{Asad_2024}, the probability of packet loss due to Doppler shift for an arbitrary packet of an arbitrary ${\rm ED}_n$ with transmission time of $t_{{\rm s},n}$ is equivalent to ${\rm Pr}\{\xi_n(t_{{\rm s},n})\geq 0.25\times {\rm BW}\}$. 

For the derivation of PLR due to Doppler shift for any packet of any ED, it should be noted that with one satellite passing, ${\rm ED}_n$ transmits $N_{\rm p}(t_{{\rm v},n},{\rm ToA}_n)$ packets. In our model, we assume that as soon as the satellite becomes visible to ${\rm ED}_n$, it starts to transmit its $N_{\rm p}(t_{{\rm v},n},{\rm ToA}_n)$ packets consecutively. Therefore, the starting time of $n_{\rm p}$th packet of ${\rm ED}_n$ is equal to $t_{{\rm s},n}+(n_{\rm P}-1){\rm ToA}_n$ for $n_{\rm P}=1,2,\dots,N_{\rm p}(t_{{\rm v},n},{\rm ToA}_n)$. Therefore, the PLR due to Doppler shift must be derived using an averaging approach over all of the random variables, i.e., $t_{{\rm s},n}$, ${\rm ToA}_n$, $N_{\rm p}(t_{{{\rm v},n}},{\rm ToA}_n)$, and $t_{{\rm v},n}$. This can be expressed as: 
\begin{IEEEeqnarray}{rCl}
	\label{eds}
	\mathcal{P}(\mathcal{E}_{\rm DS})&=&\mathbb{E}_{{\rm ToA}_n,t_{{\rm s},n},t_{{\rm v},n},N_{\rm P}}\Bigg{\{}{\rm Pr}\Big{\{}\xi_n(t_{{\rm s},n}+(n_{\rm P}-1){\rm ToA}_n)\nonumber\\
	&&\geq \frac{\rm BW}{4}
	\Big{|}{\rm ToA}_n,t_{{\rm s},n},t_{{\rm v},n},N_{\rm P}\Big{\}}\Bigg{\}}.
\end{IEEEeqnarray}
To derive the above probability analytically, we present the following lemmas. 
\begin{lemma}
	\label{lemmCDF}
 The cumulative distribution function (CDF) of random variable $\Xi$ representing the Doppler shift magnitude of ${\rm ED}_{n}$ at time $t$, $\xi_{n}(t)$, can be written as (\ref{CDF_delt}) at the top of the next page,	
\begin{figure*}
	\normalsize
\begin{IEEEeqnarray}{rCl}
	\label{CDF_delt}
	{\rm Pr}\left\{\xi_n(t)\leq f_{\rm th}\right\}&=&F_{\Xi}\left[\xi_n(t),f_{\rm th}\right]=\int_{-r}^r f_X(x)\Big{[}1+F_{Y|X}(-e+\sqrt{R_2}|x)-F_{Y|X}(-e-\sqrt{R_2}|x)+F_{Y|X}(-e-\sqrt{R_1}|x)\nonumber\\
	&&-F_{Y|X}(-e+\sqrt{R_1}|x)\Big{]}{\rm d}x,
\end{IEEEeqnarray}
\hrulefill
\vspace
*
{4pt}
\end{figure*}
where $F_{Y|X}(y|x)$ is the CDF of random variable $Y$ representing the $y$-coordinate conditioned on random variable $X$ representing the $x$-coordinate and can be expressed as
\begin{IEEEeqnarray}{rCl}
	\label{CDF_y}
	F_{Y|X}(y|x) = \frac{y-y_{\rm min}}{2y_c(0)+2\sqrt{r^2-x^2}},
\end{IEEEeqnarray}
with $y_{\rm min}=-y_c(0)-\sqrt{r^2-x^2}$. Moreover, $f_X(x)$ is the probability density function (PDF) of random variable $X$ representing $x$-coordinate which can be expressed as:
\begin{IEEEeqnarray}{rCl}
	\label{PDF_x}
	f_X(x)=\frac{2y_c(0)+2\sqrt{r^2-x^2}}{\underbrace{4ry_c(0)+\pi r^2}_{\text{total area of the coverage region}}},
\end{IEEEeqnarray}
and we have:
\begin{IEEEeqnarray}{rCl}
	\label{e}
	e = vt-y_c(0),
\end{IEEEeqnarray}
\begin{IEEEeqnarray}{rCl}
	\label{R1}
	R_1 = \frac{(a_4a_5-a_2)-\sqrt{(a_2-a_4a_5)^2+4a_1a_3a_5}}{2a_1},
\end{IEEEeqnarray}
\begin{IEEEeqnarray}{rCl}
	\label{R2}
	R_2 = \frac{(a_4a_5-a_2)+\sqrt{(a_2-a_4a_5)^2+4a_1a_3a_5}}{2a_1},
\end{IEEEeqnarray}
where $a_1=-1/(4R^4)$, $a_2=1/R^2-x^2/(2R^2)$, $a_3=1+K^2-K(2-x^2/R^2)$, $a_4=K/R^2$, and $a_5=(f_{\rm th}/\rho)^2$. 
\end{lemma}
\begin{proof}
See Appendix A.
\end{proof}
\begin{lemma}
	\label{lemmPDF}
Assuming that the starting time of an arbitrary packet from ${\rm ED}_n$ is denoted by the random variable $T_{\rm s}$ representing $t_{{\rm s},n}$. Moreover, the satellite visibility duration for the same ED is represented by another random variable $T_{\rm v}$ representing $t_{{\rm v},n}$. It is straightforward to show that for the same ED, these two random variables are not independent. Hence, the joint PDF of $T_{\rm s}$ and $T_{\rm v}$ can be expressed as:
\begin{IEEEeqnarray}{rCl}
	\label{joint_PDF}
	f_{T_{\rm v},T_{\rm s}}&&(t_{{\rm v},n},t_{{\rm s},n})=\frac{2v\sqrt{r^2-\frac{v^2t_{{\rm v},n}^2}{4}}}{4ry_c(0)+\pi r^2}\delta(t_{\rm s})\nonumber\\
	&&+\frac{v^3t_{{\rm v},n}}{2(4ry_c(0)+\pi r^2)\sqrt{r^2-\frac{v^2t_{{\rm v},n}^2}{4}}}\nonumber\\
	&&+\frac{2v\sqrt{r^2-\frac{v^2t_{{\rm v},n}^2}{4}}}{4ry_c(0)+\pi r^2}\delta\left(t_{{\rm s},n}+t_{{\rm v},n}-\frac{2y_c(0)}{v}\right),\nonumber\\
\end{IEEEeqnarray}
for $t_{{\rm v},n}\in[0,y_c(0)/v]$ and $t_{{\rm s},n}\in[0,2y_c(0)/v-t_{{\rm v},n}]$.
\end{lemma}
\begin{proof}
See Appendix B.
\end{proof}

Considering Lemma \ref{lemmCDF}, we obtained the expression for ${\rm Pr}\{\xi_n(t)\leq f_{\rm th}\}$ as in (\ref{CDF_delt}). So, we can simply substitute $t$ with $t_{{\rm s},n}+(n_{\rm P}-1){\rm ToA}_n$ and $f_{\rm th}$ with ${\rm BW}/4$ as required to derive the expression for (\ref{eds}). Also, in Lemma \ref{lemmPDF}, we obtained the joint PDF of two random variables $T_{\rm s}$ and $T_{\rm v}$ as $f_{T_{\rm v},T_{\rm s}}(t_{{\rm v},n},t_{{\rm s},n})$ in (\ref{joint_PDF}). This will help us in taking the expectation over $t_{{\rm v},n}$ and $t_{{\rm s},n}$ in (\ref{eds}). Considering these expressions, PLR due to Doppler shift, i.e., $\mathcal{P}(\mathcal{E}_{\rm DS})$ in (\ref{eds}) can be rewritten as (\ref{PLR_Dopp_shift}) at the top of the next page 
\begin{figure*}
	\normalsize
	\begin{IEEEeqnarray}{rCl}
		\label{PLR_Dopp_shift}
		\mathcal{P}(\mathcal{E}_{\rm DS})&=&\frac{1}{N_{\rm ToA}}\sum_{i=1}^{N_{\rm ToA}}\int_{100 {\rm ToA}_i}^{\frac{y_c(0)}{v}}\frac{1}{N_{\rm p}(t_{{\rm v}},{\rm ToA}_i)}\sum_{j=0}^{N_{\rm p}(t_{{\rm v}},{\rm ToA}_i)-1}\int_0^{\frac{2y_c(0)}{v}-t_{\rm v}}f_{T_{\rm v},T_{\rm s}}(t_{\rm v},t_{\rm s})\left[1-F_\Xi\left(\xi(t_{\rm s}+j {\rm ToA}_i),0.25{\rm BW}\right)\right]\nonumber\\
		&&\times{\rm d}t_{\rm s}{\rm d}t_{\rm v},
	\end{IEEEeqnarray}
	\hrulefill
	\vspace
	*
	{4pt}
\end{figure*}
where $N_{\rm ToA}$ is the number of all possible values of $\rm ToA$. It can be observed that the term $1-F_{\Xi}(\xi(t_{\rm s}+j{\rm ToA}_i),0.25{\rm BW})$ within the equation (\ref{PLR_Dopp_shift}) depends on the $\rm BW$. Furthermore, considering (\ref{CDF_delt}), we can write $F_{\Xi}(\xi(t_{\rm s}+j{\rm ToA}_i),0.25{\rm BW})={\rm Pr}\{\xi(t_{\rm s}+j{\rm ToA}_i))\leq 0.25 {\rm BW}\}$. Since increasing $\rm BW$ enlarges the threshold $0.25\rm BW$, it can be concluded that $F_{\Xi}(\xi(t_{\rm s}+j{\rm ToA}_i),0.25{\rm BW})$ is a non-decreasing function of $\rm BW$, and $\mathcal{P}(\mathcal{E}_{\rm DS})$ is a decreasing function of $\rm BW$ because of the $1-F_{\Xi}(\xi(t_{\rm s}+j{\rm ToA}_i),0.25{\rm BW})$ term. It can be inferred that for larger values of $\rm BW$, the packet loss probability caused by Doppler shift is reduced. It should be noted that for packet loss to happen, we must have $N_{\rm p}(t_{{\rm v},n},{\rm ToA}_n)\geq 1$ which is the reason that the outer integral lower limit is $100 {\rm ToA}_i$ instead of zero (value $100$ comes from the inverse of the duty cycle $1\%$). It is also noteworthy that the PLR due to Doppler shift expressed in (\ref{PLR_Dopp_shift}) does not have an analytical solution and can be solved numerically. 
\subsection{Packet loss due to Doppler rate}
According to \cite{Asad_2024}, the probability of packet loss due to Doppler rate for an arbitrary packet of an arbitrary ${\rm ED}_n$ with transmission time of $t_{{\rm s},n}$ is equivalent to ${\rm Pr}\{|\xi(t_{{\rm s},n})-\xi(t_{{\rm s},n}+{\rm ToA}_n)|\geq {\rm BW}/(3\times 2 ^{\rm SF})\}$.

Hence, similar to (\ref{eds}), $\mathcal{P}(\mathcal{E}_{DR})$ can be expressed as:
\begin{IEEEeqnarray}{rCl}
	\label{edr}
	\mathcal{P}&&(\mathcal{E}_{\rm DR})=\mathbb{E}_{{\rm ToA}_n,t_{{\rm s},n},t_{{\rm v},n},N_{\rm P}}\Bigg{\{}{\rm Pr}\Big{\{}\Big{|}\xi_n(t_{{\rm v},n}+(n_{\rm P}-1){\rm ToA}_n\nonumber\\
	&&-\xi_n(t_{{\rm v},n}+n_{\rm P}{\rm ToA}_n)\Big{|}\geq \frac{\rm BW}{3\times 2^{\rm SF}_n}\Big{|}{\rm ToA}_n,t_{{\rm s},n},t_{{\rm v},n},N_{\rm P}\Big{\}}\Bigg{\}}.\nonumber\\
\end{IEEEeqnarray}
Using the same approach used in (\ref{Prob_CDF_Dopp_xy}), we can start the derivation of (\ref{edr}) by rewriting the Doppler function in terms of $x$ and $y$ and write the general equation of (\ref{Dopp_rate_xy}) for any $t$ and $\rm ToA$ at the top of the next page. 
\begin{figure*}
	\normalsize
	\begin{IEEEeqnarray}{rCl}
		\label{Dopp_rate_xy}
		{\rm Pr}\left\{|\xi(t)-\xi(t+{\rm ToA})|\geq f_{\rm th}\right\}&=&{\rm Pr}\Bigg{\{}\Bigg{|}\rho\sqrt{\frac{{\cos^2\left[2\sin^{-1}\left(\frac{|x|}{2R}\right)\right]-\cos^2\left[2\sin^{-1}\left[\frac{x^2+(y-y_c(0)+vt)^2}{2R}\right]\right]}}{1+K^2-2K\cos\left[2\sin^{-1}\left[\frac{x^2+(y-y_c(0)+vt)^2}{2R}\right]\right]}}\nonumber\\
		&&-\rho\sqrt{\frac{{\cos^2\left[2\sin^{-1}\left(\frac{|x|}{2R}\right)\right]-\cos^2\left[2\sin^{-1}\left[\frac{x^2+(y-y_c(0)+v(t+{\rm ToA}))^2}{2R}\right]\right]}}{1+K^2-2K\cos\left[2\sin^{-1}\left[\frac{x^2+(y-y_c(0)+v(t+{\rm ToA}))^2}{2R}\right]\right]}}\Bigg{|}\geq f_{\rm th}\Bigg{\}}.
	\end{IEEEeqnarray}
	\hrulefill
	\vspace
	*
	{4pt}
\end{figure*}
Here, we follow the same procedure as in Appendix \ref{appA}, and apply the marginalization over $x$-coordinate and then averaging with respect to $f_X(x)$. But the issue is that the inequality resulting from mathematical simplification of (\ref{Dopp_rate_xy}) is $16$th-degree for which the roots must be obtained numerically. So, we can rewrite (\ref{Dopp_rate_xy}) as:
\begin{IEEEeqnarray}{rCl}
	\label{simple_Dopp_rate_xy}
	{\rm Pr}&&\left\{|\xi(t)-\xi(t+{\rm ToA})|\geq f_{\rm th}\right\}=\nonumber\\
	&&\int_{-r}^rf_X(x){\rm Pr}\Big{\{}\iota(x=\kappa,y)\geq 0|x=\kappa\Big{\}}{\rm d}x,
\end{IEEEeqnarray}
where $\iota(x,y)$ is a $16$th-degree polynomial expressed as (\ref{iota}) at the top of the next page. 
\begin{figure*}
	\normalsize
	\begin{IEEEeqnarray}{rCl}
		\label{iota}
		\iota(x,y)&=&\Bigg{|}\rho\sqrt{\frac{{\cos^2\left[2\sin^{-1}\left(\frac{|x|}{2R}\right)\right]-\cos^2\left[2\sin^{-1}\left[\frac{x^2+(y-y_c(0)+vt)^2}{2R}\right]\right]}}{1+K^2-2K\cos\left[2\sin^{-1}\left[\frac{x^2+(y-y_c(0)+vt)^2}{2R}\right]\right]}}\nonumber\\
		&&-\rho\sqrt{\frac{{\cos^2\left[2\sin^{-1}\left(\frac{|x|}{2R}\right)\right]-\cos^2\left[2\sin^{-1}\left[\frac{x^2+(y-y_c(0)+v(t+{\rm ToA}))^2}{2R}\right]\right]}}{1+K^2-2K\cos\left[2\sin^{-1}\left[\frac{x^2+(y-y_c(0)+v(t+{\rm ToA}))^2}{2R}\right]\right]}}\Bigg{|}-f_{\rm th}.
	\end{IEEEeqnarray}
	\hrulefill
	\vspace
	*
	{4pt}
\end{figure*}
Based on this, we define the following region in terms of $y$ for which $\iota(x=\kappa,y)\geq0$ as 
\begin{IEEEeqnarray}{rCl}
	\label{region_y}
	\mathcal{Y}(\kappa)=\{y\in\mathbb{R}:\iota(\kappa,y)\geq 0\}=\bigcup_{m=1}^{M(\kappa)}\left[y_m^{\rm lower}(\kappa),y_m^{\rm upper}(\kappa)\right],\nonumber\\
\end{IEEEeqnarray}
where $M(\kappa)$ is the number of valid intervals for which $\iota(x=\kappa,y)\geq0$ for the given $x=\kappa$. Obviously, the endpoints $y_m^{\rm lower}(\kappa)$ and $y_m^{\rm upper}(\kappa)$ are the numerically obtained roots. Consequently, we can express the probability in (\ref{simple_Dopp_rate_xy}) as (\ref{simple_Dopp_iota}) at the top of the next page.
\begin{figure*}
	\normalsize
	\begin{IEEEeqnarray}{rCl}
		\label{simple_Dopp_iota}
		{\rm Pr}\left\{|\xi(t)-\xi(t+{\rm ToA})|\geq f_{\rm th}\right\}=\Lambda(t,{\rm ToA},f_{\rm th})=\int_{-r}^r f_X(x)\sum_{m=1}^{M(x)}\left[F_{Y|X}(y_m^{\rm upper}(x)|x)-F_{Y|X}(y_m^{\rm lower}(x)|x)\right]{\rm d}x.
	\end{IEEEeqnarray}
	\hrulefill
	\vspace
	*
	{4pt}
\end{figure*}
Finally, the total PLR due to Doppler rate can be expressed as (\ref{PLR_Dopp_rate}) at the top of the next page. Similar to the discussion provided for $\mathcal{P}(\mathcal{E}_{\rm DS})$, and considering (\ref{simple_Dopp_iota}) and (\ref{PLR_Dopp_rate}), it can be inferred that $\mathcal{P}(\mathcal{E}_{\rm DR})$ is a decreasing function of $\rm BW$ and an increasing function of $\rm SF$. 
\begin{figure*}
	\normalsize
	\begin{IEEEeqnarray}{rCl}
		\label{PLR_Dopp_rate}
		\mathcal{P}(\mathcal{E}_{\rm DR})&=&\frac{1}{N_{\rm ToA}}\sum_{i=1}^{N_{\rm ToA}}\int_{100 {\rm ToA}_i}^{\frac{y_c(0)}{v}}\frac{1}{N_{\rm p}(t_{{\rm v}},{\rm ToA}_i)}\sum_{j=0}^{N_{\rm p}(t_{{\rm v}},{\rm ToA}_i)-1}\int_0^{\frac{2y_c(0)}{v}-t_{\rm v}}f_{T_{\rm v},T_{\rm s}}(t_{\rm v},t_{\rm s})\Lambda\left(t_{\rm s}+j{\rm ToA}_i,{\rm ToA}_i,\frac{\rm BW}{3\times 2^{\rm SF}}\right){\rm d}t_{\rm s}{\rm d}t_{\rm v},\nonumber\\
	\end{IEEEeqnarray}
	\hrulefill
	\vspace
	*
	{4pt}
\end{figure*}

\subsection{Packet loss due to collision}
Packet loss due to collision is defined as the event that at least one packet interferes with the desired packet in both frequency and time domains. Therefore, we define the following packet vulnerabilities in frequency and time:
\begin{itemize}
	\item Frequency vulnerability: For a packet transmitted by an ED, e.g., ${\rm ED}_n$, to be vulnerable to another packet from an interfering ED, e.g., ${\rm ED}_m$, in the frequency domain simply means that both ${\rm ED}_n$ and ${\rm ED}_m$ select the same channel out of the $N_{\rm CH}$ available channels for packet transmission, i.e., ${\rm CH}_n={\rm CH}_m$. 
	\item Time vulnerability: Suppose that the starting time of a packet transmission by ${\rm ED}_n$ is $\tau_n$ and the ending time is $\tau_n+{\rm ToA}_n$ where $\tau_n$ can be $t_{{\rm s},n}$, $t_{{\rm s},n}+{\rm ToA}_n$, $\dots$, or $t_{{\rm s},n}+(N_{\rm p}(t_{{\rm v},n},{\rm ToA}_n)-1){\rm ToA}_n$. The vulnerability duration in the time domain for this packet to be overlapped by another packet from an interfering device ${\rm ED}_m$ with starting time $\tau_m$ and ending time $\tau_m+{\rm ToA}_m$ can be expressed as:
	
\begin{IEEEeqnarray}{rCl}
	\label{t_vul}
	\mathcal{T}_{n,\rm vul} = \left\{t:\tau_n-{\rm ToA}_m\leq t \leq \tau_n+{\rm ToA}_n\right\}.
\end{IEEEeqnarray}
	Therefore, if $\tau_m\in{\mathcal{T}_{n,{\rm vul}}}$, the packet transmitted by ${\rm ED}_m$ collides with the packet transmitted by ${\rm ED}_n$ in the time domain. 
\end{itemize}

Now, assume a specific packet transmitted by ${\rm ED}_n$ with the starting time of $\tau_n$ and ending time of $\tau_n+{\rm ToA}_n$. The probability that at least one packet from the other $N-1$ EDs overlap with this packet in time and frequency can be written as:
\begin{IEEEeqnarray}{rCl}
	\label{P_Col_tagged}
	P_{\rm col}(\tau_n,{\rm ToA}_n) = 1-\left[1-q(\tau_n,{\rm ToA}_n)\right]^{N-1}
\end{IEEEeqnarray}
where $q(\tau_n,{\rm ToA}_n)$ can be represented as (\ref{q_tau}) at the top of the next page. The rationale behind (\ref{q_tau}) is that the vulnerable period associated with the desired packet is a function of the transmission start time and ToA of the interfering ED. Consequently, the collision probability must be averaged over all possible realizations of these parameters for a single interfering ED. In addition, the factor $1/N_{\rm CH}$ accounts for the probability that the interfering ED transmits on the same channel as ${\rm ED}_n$. Also, $\mathbbm{1}\{\mathcal{C}\}$ is the indicator function where $\mathbbm{1}\{\mathcal{C}\}=1$ if $\mathcal{C}$ is true and $\mathbbm{1}\{\mathcal{C}\}=0$ otherwise.
\begin{figure*}
	\normalsize
	\begin{IEEEeqnarray}{rCl}
		\label{q_tau}
		q(\tau_n,{\rm ToA}_n)&=&\frac{1}{N_{\rm CH}}\frac{1}{N_{\rm ToA}}\sum_{j=1}^{N_{\rm ToA}}\int_{100 {\rm ToA}_j}^{\frac{y_c(0)}{v}}\int_0^{\frac{2y_c(0)}{v}-t_{\rm v}}f_{T_{\rm v},T_{\rm s}}(t_{\rm v},t_{\rm s})\nonumber\\
		&&\times\mathbbm{1}\Big{\{}\exists k \in \left\{0,\dots,N_p(t_{{\rm v},n},{\rm ToA}_n)\right\}:	t_{\rm s}+k{\rm ToA}_j-{\rm ToA}_n\leq \tau_n \leq t_{\rm s}+(k+1){\rm ToA}_j\Big{\}}{\rm d}t_{\rm s}{\rm d}t_{\rm v},
	\end{IEEEeqnarray}
	\hrulefill
	\vspace
	*
	{4pt}
\end{figure*}
Eventually, the total PLR due to collision can be obtained as (\ref{PLR_col}) at the top of the next page. From (\ref{P_Col_tagged}), (\ref{q_tau}), and (\ref{PLR_col}), it can be observed that the PLR due to collisions increases with the duration of the packet vulnerable period. This is because a longer vulnerable period increases the likelihood of time-domain overlap with transmissions from other EDs. Since the vulnerable period is directly proportional to the $\rm ToA$, EDs employing larger ToA values are more susceptible to collisions. According to (\ref{ToA}), larger $\rm ToA$ values are associated with higher values of $\rm SF$ and lower values of $\rm CR$ (more robust coding configurations), implying that such EDs experience a higher collision-induced PLR.
\begin{figure*}
	\normalsize
	\begin{IEEEeqnarray}{rCl}
	\label{PLR_col}
	\mathcal{P}(\mathcal{E}_{\rm Col})&=&\mathbb{E}_{{\rm ToA},t_{\rm s},t_{\rm v},N_{\rm P}}\Bigg{\{}P_{\rm col}\left[t_{\rm s}+(n_{\rm P}-1){\rm ToA},{\rm ToA}\right]\Big{|}{\rm ToA},t_{\rm s},t_{\rm v},N_{\rm P}\Big{\}}\Bigg{\}}=\nonumber\\
	&&\frac{1}{N_{\rm ToA}}\sum_{i=1}^{N_{\rm ToA}}\int_{100 {\rm ToA}_i}^{\frac{y_c(0)}{v}}\frac{1}{N_{\rm p}(t_{{\rm v}},{\rm ToA}_i)}\sum_{j=0}^{N_{\rm p}(t_{{\rm v}},{\rm ToA}_i)-1}\int_0^{\frac{2y_c(0)}{v}-t_{\rm v}}f_{T_{\rm v},T_{\rm s}}(t_{\rm v},t_{\rm s})P_{\rm col}\left[t_{\rm s}+j{\rm ToA}_i,{\rm ToA}_i\right]{\rm d}t_{\rm s}{\rm d}t_{\rm v}.
	\end{IEEEeqnarray}
	\hrulefill
	\vspace
	*
	{4pt}
\end{figure*}

Finally, based on the previous derivations, the total packet loss ratio can be expressed as:
\begin{IEEEeqnarray}{rCl}
	\label{PLR_tot}
	{\rm PLR} &=& \mathcal{P}(\mathcal{E}_{\rm DS})+[1-\mathcal{P}(\mathcal{E}_{\rm DS})]\mathcal{P}(\mathcal{E}_{\rm DR})\nonumber\\
	&&+[1-\mathcal{P}(\mathcal{E}_{\rm DS})][1-\mathcal{P}(\mathcal{E}_{\rm DR})]\mathcal{P}(\mathcal{E}_{\rm Col}).
\end{IEEEeqnarray}
Based on the preceding discussions on $\mathcal{P}(\mathcal{E}_{\rm DS})$ and $\mathcal{P}(\mathcal{E}_{\rm DR})$ (both are decreasing functions of $\rm BW$), it can be inferred that, for sufficiently large values of $\rm BW$, the effects of both Doppler shift and Doppler rate become negligible. In this case, the overall PLR is dominated by packet collisions and approaches $\mathcal{P}(\mathcal{E}_{\rm Col})$. However, for the practical bandwidth values employed in LoRaWAN systems, the simulation results demonstrate that only the effect of Doppler shift can be safely neglected, whereas Doppler rate remains a significant impairment and contributes non-negligibly to the total PLR.
\section{Reinforcement Learning-based Doppler-aware Resource Allocation}
\label{RL}
The goal here is to improve the PLR performance of the presented DtS-IoT LoRaWAN system model by designing a resource allocation approach that assigns ${\rm SF}$, ${\rm BW}$, ${\rm CR}$, and packet transmission starting times for each ED effectively. This can be formulated as an optimization problem as (for $i=1,2,\dots,N$ and $k_i=1,\ldots,N_{{\rm p}}(t_{v,i},{\rm ToA}_i)$):
\begin{IEEEeqnarray}{rCl}
	\label{Opt}
	\mathcal{P}_{\rm plr}&:&
	\min_{\mathbf{SF},\mathbf{BW},\mathbf{CR},\mathbf{T}^{(\rm p)}}
	{\rm PLR}
	\nonumber\\
	&{\rm s.t.\;\;}&
	{\rm SF}_i\in\{7,8,9,10\},
	\quad \forall i,
	\nonumber\\
	&&
	{\rm BW}_i\in\{125,250,500\}\ {\rm kHz},
	\quad \forall i,
	\nonumber\\
	&&
	{\rm CR}_i\in\{4/5,4/6,4/7,4/8\},
	\quad \forall i,
	\nonumber\\
	&&
	t_{{\rm s},i}+(k_i-1){\rm ToA}_i
	\leq \tau_{i,k_i}
	\nonumber\\
	&&\hspace{2mm}\leq
	t_{{\rm e},i}
	-\left(N_{{\rm p}}(t_{v,i},{\rm ToA}_i)-k_i+1\right){\rm ToA}_i,\quad \forall i,k_i,
	\nonumber\\
	&&
	\tau_{i,k_i+1}
	\geq \tau_{i,k_i}+{\rm ToA}_i,
	\quad \forall i,k_i,
\end{IEEEeqnarray}
 where ${\rm \mathbf{SF}}=[{\rm SF}_1,{\rm SF}_2,\dots,{\rm SF}_N]^\top$ is the vector of assigned $\rm SF$ values for all $N$ EDs, ${\rm \mathbf{BW}}=[{\rm BW}_1,{\rm BW}_2,\dots,{\rm BW}_N]^\top$ is the vector of assigned $\rm BW$ values for all $N$ EDs, and ${\rm \mathbf{CR}}=[{\rm CR}_1,{\rm CR}_2,\dots,{\rm CR}_N]^\top$ is the vector of assigned $\rm CR$ values for all $N$ EDs. Also, $\mathbf{T}^{(\rm p)}=[\mathbf{t}^{({\rm p})}_1,\mathbf{t}^{({\rm p})}_2,\dots,\mathbf{t}^{({\rm p})}_N]^\top$ is a ragged array consisting of the packet transmission times for each ED in which $\mathbf{t}_i^{({\rm p})}=[\tau_{i,1},\tau_{i,2},\dots,\tau_{i,N_{{\rm p}}(t_{v,i},{\rm ToA}_i)}]^\top$ is the vector of starting transmission times for packets of ${\rm ED}_i$.
 
 From (\ref{Opt}), it can be seen that the problem of joint resource allocation for every ED in the network to minimize the total PLR is a mixed-integer, non-convex problem. The decision variables include three discrete assignment vectors ($\mathbf{SF}$, $\mathbf{BW}$, $\mathbf{CR}$)
 and the continuous ragged array $\mathbf{T}^{(\rm p)}$ of per-ED packet start times. Also, the objective function PLR couples every ED's decisions to every other ED's decisions through the term $\mathcal{P}(\mathcal{E}_{\rm col})$ as presented in (\ref{PLR_tot}). On the other hand, for every network realization, the optimization problem ${\mathcal{P}_{\rm plr}}$ needs to be solved again. The set of visible EDs, along with their positions, visibility intervals, and related collision conditions, are constantly changing as the satellite moves around its orbit. As a result, an instance-specific optimization technique whose computational complexity increases super-linearly with the number of EDs, $N$, is not scalable for large networks and might not be appropriate for on-board resource allocation under time constraints. For this reason, we adopt a reinforcement learning (RL) approach which is well suited for large-scale resource allocation problems in wireless communications \cite{Surv_ML_RA}. The principal advantage of applying RL resource allocation to our problem here is not that it necessarily
 achieves a lower $\mathrm{PLR}$ than a well-tuned per-instance solver of the optimization problem $\mathcal{P}_{\rm plr}$, but that it produces assignments at constant amortized cost and generalizes to network sizes far beyond those on which it was trained. Accordingly, in the following, we present the RL approach used for resource allocation in our system model.
 
 Fig. \ref{RL_SAT} shows the block diagram of the exploited RL network. The following are the descriptions of the elements in the architecture:
 \begin{figure}[t!]
 	\centering
 	\includegraphics[width=\linewidth]{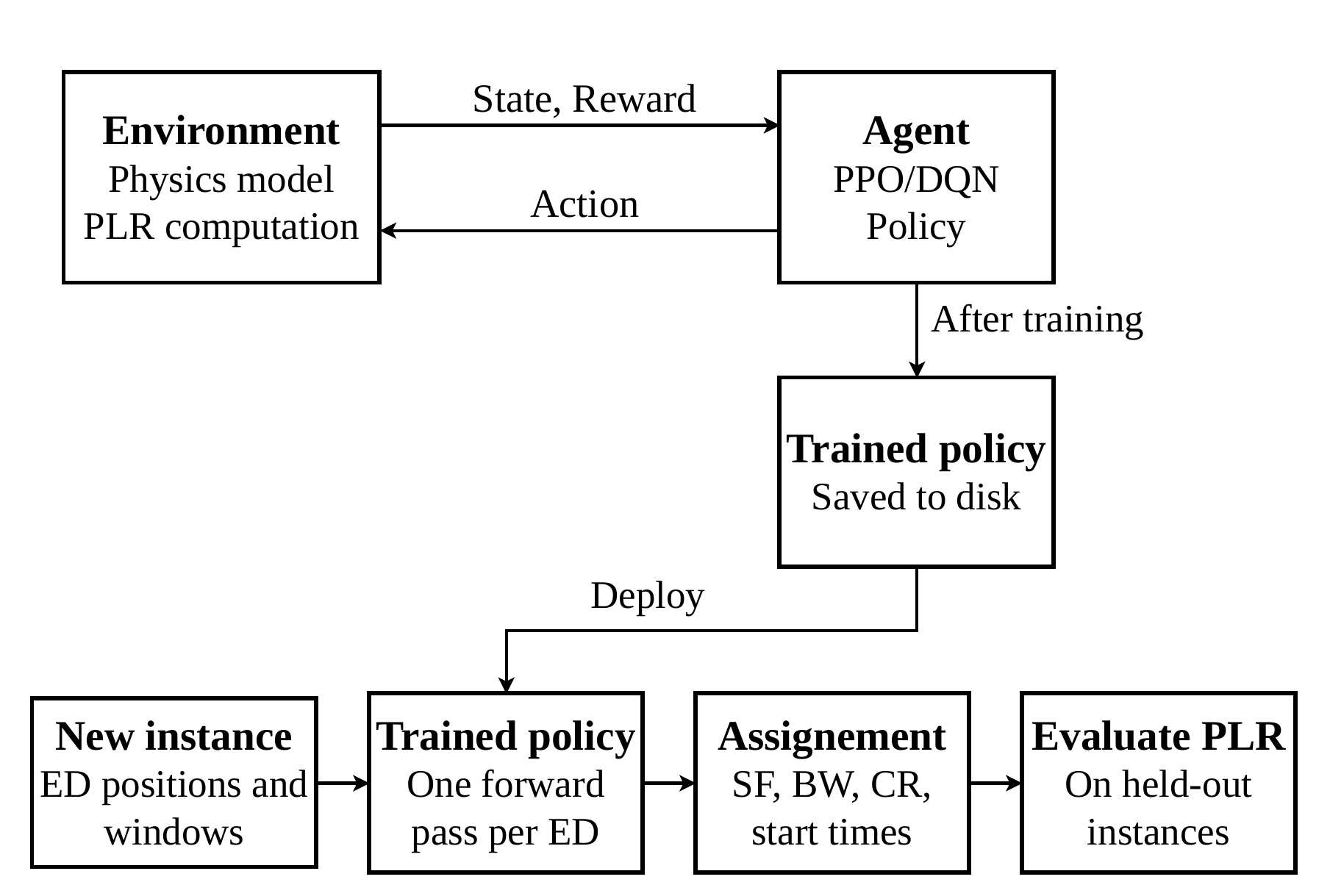}
 	\caption{Block diagram of the exploited RL architecture.}
 	\label{RL_SAT}
 \end{figure}
 
 \subsubsection{\textbf{Environment block}} The environment includes the full physical model of the presented DtS-IoT LoRaWAN scenario, i.e., ED positions and per-ED visibility windows $[t_{{\rm s},i},t_{{\rm e},i}]$ generated from the satellite pass geometry, and the Doppler shift and Doppler rate profiles seen by each ED throughout its window. Also, given any candidate transmission schedule for a selected ${\rm ED}_i$, the environment computes the resulting packet $k_i$ outcome with starting time $\tau_{k_i}$ and ending time $\tau_{k_i}+{\rm ToA}_i$ under the three loss mechanisms of Section~\ref{plr} and returns the following per-packet PLR contribution:
 \begin{IEEEeqnarray}{rCl}
 	\label{PLR_rew}
{\rm PLR}^{(\rm pp)}&&(\tau_{k_i},{\rm ToA}_i)=1-\mathbbm{1}\left\{\xi(\tau_{k_i})\leq 0.25\times{\rm BW}_i\right\}\nonumber\\
&&\times\mathbbm{1}\left\{\Big{|}\xi(\tau_{k_i})-\xi(\tau_{k_i}+{\rm ToA}_i)\Big{|}\leq \frac{{\rm BW}_i}{3\times 2^{{\rm SF}_i}}\right\}\nonumber\\
&&\times \left[1-q(\tau_{k_i},{\rm ToA}_i)\right],
 \end{IEEEeqnarray}
 where $q(\tau_{k_i},{\rm ToA}_i)$ denotes the per-packet loss ratio due to collision and can be calculated using (\ref{q_tau}).

It should be noted that (\ref{q_tau}) assumes that each ED starts transmitting upon satellite visibility and transmits its packets consecutively. This assumption does not hold in the proposed RL framework, where transmissions are scheduled according to the learned policy. Consequently, the exact collision PLR cannot be derived in closed form, as it depends on the joint, policy-dependent transmission decisions of all EDs. Such interactions make the collision process analytically intractable, motivating the use of surrogate or approximate analytical models to guide RL-based resource allocation.
 
 Nevertheless, we consider $q(\tau_{k_i},{\rm ToA}i)$ as the collision component of the per-packet PLR model, and consequently as part of the reward function, since it captures the dominant relationship between packet duration and collision probability. Specifically, the packet duration is uniquely determined by the transmission parameters $\rm SF$, $\rm CR$, and $\rm BW$, and longer packets occupy the channel for a larger fraction of time, thereby increasing their likelihood of overlapping with transmissions from other EDs. Therefore, $q(\tau_{k_i},{\rm ToA}_i)$ serves as a computationally efficient surrogate for the true collision-induced PLR. Although it does not explicitly account for the policy-dependent transmission schedules learned by the RL agent, it preserves the fundamental dependence of collision probability on packet duration while avoiding the prohibitive complexity of deriving the exact per-packet PLR due to collision under adaptive scheduling.
 \subsubsection{\textbf{State, Action, and Reward}}   
The RL formulation of $\mathcal{P}_{\rm plr}$ is defined by the tuple $({\rm State}, {\rm Action}, {\rm Reward})$ consisting of the state space $\rm State$,
the action space $\rm Action$, and the reward function $\rm Reward$ defined as follows:
\begin{itemize}
	\item State: For ${\rm ED}_i$ the state ${\rm state}_i \in \mathbb{R}^{3}$
	observed by the agent is
	\begin{IEEEeqnarray}{rCl}
		\label{state}
{\rm state}_i=\Big[\frac{t_{{\rm v},i}}{\max_j t_{{\rm v},j}}, \frac{t_{{\rm s},i}}{T_{\rm SAT}},\frac{L_{{\rm c},i}}{\max_j L_{{\rm c},j}}\Big]^\top,
	\end{IEEEeqnarray}
	where $L_{{\rm c},i}$ is the number of EDs sharing ${\rm ED}_i$ channel. It should be noted that the choice of ${\rm state}_i$ in (\ref{state}) is guided directly
	by the PLR expression derived in Section~\ref{plr}.
	The number of packets $N_{\rm p}(t_{v,i},{\rm ToA}_i)$ that ED~$i$
	transmits is fully determined by its visibility window duration
	$t_{v,i}$, motivating the first feature. The second feature locates the window within the pass, which affects both the Doppler profile seen by the packets and their exposure to other transmissions. Finally, the channel load $L_{{\rm c},i}$ measures congestion on the ${\rm ED}_i$'s channel, the primary driver of collision. 

	\item Action: For each ED, the agent selects an action
${\rm action}_i \in \rm Action$ consisting of a discrete
$({\rm SF},{\rm CR},{\rm BW})$ combination and a discrete first-packet
timing bin:

\begin{IEEEeqnarray}{rCl}
	\label{action}
	{\rm action}_i=\left[c_i,\; n_{t_i}\right]^\top,\;\; &&c_i\in\{0,1,\ldots,47\},\nonumber\\
	&&n_{t_i}\in\{0,1,\ldots,N_t-1\},
\end{IEEEeqnarray}

where the combo index $c_i$ selects one of the $4\times4\times3=48$
$({\rm SF},{\rm CR},{\rm BW})$ tuples and the timing bin $n_{t_i}$ discretizes the starting time of ${\rm ED}_i$'s first packet within its visibility window using $N_{t}$
uniformly spaced bins. Given $n_{t_i}$, the first packet's start time is set to
\begin{IEEEeqnarray}{rCl}
	\label{eq:tau_first}
	\tau_{i,1}&=&t_{{\rm s},i}
	+\frac{n_{t_i}}{N_t-1}\left[t_{v,i} - N_{\rm p}(t_{v,i},{\rm ToA}_i)\,{\rm ToA}_i \right],\nonumber\\
\end{IEEEeqnarray}
and the remaining $\tau_{i,k_i}$ for $k_i \geq 2$ are placed consecutively at
the spacing $\tau_{i,k_i+1}-\tau_{i,k_i}={\rm ToA}_i$, so that ${\rm ED}_i$
transmits its $N_{\rm p}$ packets back-to-back starting from $\tau_{i,1}$.
Consequently, a single action per ED fully determines that ED's transmission
schedule.
\item Reward: The reward returned by the environment after ${\rm ED}_i$'s action is the
mean success probability of the packets scheduled for that ED:
\begin{IEEEeqnarray}{rCl}
	\label{reward}
	{\rm reward}_i&=&\frac{1}{N_{\rm p}(t_{v,i},{\rm ToA}_i)}\nonumber\\
	&&\times\sum_{k_i=1}^{N_{\rm p}(t_{v,i},{\rm ToA}_i)}\left[1 - {\rm PLR}^{(\rm pp)}(\tau_{i,k_i})\right].\nonumber\\
\end{IEEEeqnarray}
Accordingly, the learning algorithm adjusts the parameters $\theta$ of the policy
$\pi_\theta$ (introduced in next subsection) to maximize the expected discounted return
\begin{IEEEeqnarray}{rCl}
	\label{objective}
	\mathcal{R}(\theta)=\mathbb{E}_{\pi_\theta}\left\{\sum_{i=1}^{N} \gamma^{\,i-1}\, {\rm reward}_i\right\},
\end{IEEEeqnarray}
where $\gamma\in(0,1]$ is the discount factor. The discount factor $\gamma$ determines how much the agent values future rewards. Scheduling ${\rm ED}_i$ affects not only its own reward ${\rm reward}_i$ but also the collision environment experienced by the coming scheduled EDs. Therefore, maximizing only the immediate reward may degrade the overall network performance. A larger $\gamma$ encourages the agent to consider the long-term impact of its decisions, while a smaller value results in a more greedy policy. In this work, we set $\gamma=0.95$, a common choice in RL corresponding to an effective planning horizon of approximately $20$ scheduling decisions \cite{RL_policy}. Finally, the optimal parameters are obtained as $\theta^\star = \arg\max_\theta \mathcal{R}(\theta)$.
\end{itemize}
\subsubsection{\textbf{Agent}}
Agent is the learning component of the workflow. It maintains a parameterized policy $\pi_\theta$ whose parameters $\theta$ are adjusted during training to maximize the objective (\ref{objective}). For each ${\rm ED}_i$, the policy gets ${\rm state}_i$ as the input defined in (\ref{state}) and outputs ${\rm action}_i = [c_i, n_{t_i}]^\top$ defined in (\ref{action}). 

In this paper, we consider two learning algorithms, i.e., PPO \cite{PPO1,PPO2} and DQN \cite{DQN}. Both are trained against the same environment, use the same state and action definitions, and receive
the same reward ${\rm reward}_i$, so that any observed performance difference between them reflects the algorithmic choice rather than a difference in problem formulation. The rationale behind these selections is as follows. 

We evaluate the proposed RL formulation using DQN and PPO as representative value-based and policy-based methods, respectively \cite{RL_book}. Both share the same environment, state, reward, and discrete action space, with each joint decision $(c_i,n_{t_i})$ mapped bijectively to one of $48N_t$ actions. PPO learns a masked categorical policy, whereas DQN estimates action $Q$-values, enabling comparison of the two approaches for DtS-IoT resource allocation.

Finally, it should be noted that a tiny feed-forward neural network with one hidden layer of $64$ units and $\tanh$ activations is used to implement the policy $\pi_\theta$. The network's output layer varies depending on the algorithm, i.e., for PPO, it produces a probability distribution over the $48 \times N_t$ joint action space; for DQN, it produces a $Q$-value for every one of the identical $48 \times N_t$ discrete actions. Once trained, evaluating $\pi_\theta$ for a single ED requires only one forward pass through
this network, so the total inference cost is $\mathcal{O}(N)$ and is dominated by matrix multiplications of size independent of $N$.

\subsubsection{\textbf{Trained Policy}}
The resulting policy parameters $\theta^\star$, which correspond to the weights of the single hidden layer policy network, are written to disk as a compact static artifact when the objective (\ref{objective}) has converged. This artifact, which is a frozen snapshot of the network rather than an active calculation, is represented by the ``trained policy'' block in the training half of the picture. This is the end of learning, and once the policy has been saved, no more changes are made to its parameters.
\subsubsection{\textbf{New instance}}
Every deployment run starts with a new instance of the $\mathcal{P}_{\rm plr}$ resource allocation problem. The number of EDs $N$, their locations inside the coverage area, and the resulting per-ED visibility windows $[t_{{\rm s},i}, t_{{\rm e},i}]$ obtained from the satellite pass geometry define an instance. So basically, different instances correspond to the different random realizations of the ED spatial distribution.
\subsubsection{\textbf{Trained policy in deployment}}
During deployment, the trained policy is loaded into memory and the $N$ EDs are scheduled sequentially. For ${\rm ED}_i$, ${\rm state}_i$ in (\ref{state}) is fed to the trained network to select ${\rm action}_i$ from (\ref{action}). While actions are sampled probabilistically during training to encourage exploration, deterministic selection is employed during deployment. Since scheduling each ED requires only a single forward pass through the trained network, an entire satellite pass can be scheduled in under one second for the values of $N$ considered in this work.
\subsubsection{\textbf{Assignment}}
The complete network-wide assignment $(\mathbf{SF}, \mathbf{BW}, \mathbf{CR}, \mathbf{T}^{(\rm p)})$ that represents a potential solution to $\mathcal{P}_{\rm plr}$ in (\ref{Opt}) is obtained by concatenating the per-ED actions generated across all $i \in\{1, 2, \ldots, N\}$. The vectors $\mathbf{SF}$, $\mathbf{CR}$, $\mathbf{BW}$ are obtained by mapping each combo index $c_i$ to its corresponding $({\rm SF}, {\rm CR},{\rm BW})$ tuple and each ED's first packet is placed at $\tau_{i,1}$ of \eqref{eq:tau_first} and its remaining packets at the duty-cycle-compliant spacing indicated by \eqref{Opt}. This assignment is sent straight to the next-described evaluation block without any additional optimization or refinement.
\subsubsection{\textbf{Evaluate}}
The obtained assignment is scored by the same physical model of Section~\ref{plr}, but with the collision outcome evaluated directly from the resulting schedule. For every packet, the Doppler shift and Doppler rate conditions are checked at its actual transmission time, and a collision is declared whenever the packet overlaps in time with a packet of another ED transmitting on the same channel. Denoting the resulting per-packet loss indicator by ${\rm PLR}^{(\rm pp,ov)}(\tau_{k_i},{\rm ToA}_i)\in\{0,1\}$, the network-wide packet loss ratio for the evaluation stage is calculated as
\begin{IEEEeqnarray}{rCl}
	\label{eq:plr_eval}
	\mathrm{PLR}^{(\rm RL)}=\frac{\sum_{i=1}^{N}\sum_{k_i=1}^{N_{\rm p}(t_{v,i},{\rm ToA}_i)}
		{\rm PLR}^{(\rm pp,ov)}(\tau_{k_i},{\rm ToA}_i)}{\sum_{i=1}^{N} N_{\rm p}(t_{v,i},{\rm ToA}_i)},\nonumber\\
\end{IEEEeqnarray}
i.e. the total number of lost packets divided by the total number of transmitted packets across the network. Thus the surrogate is used only to train the policy, while performance is reported using the exact overlap-based collision model. So accordingly, we can state that the RL agent learned the policy using the derived ${\rm PLR}^{(\rm pp)}(\tau_{k_i},{\rm ToA}_i)$ at the training stage, and for evaluating scenarios, we calculate the actual PLR by counting all of the lost packets throughout the network to evaluate the performance of the learned policy.

It should be noted that the proposed RL policy is implemented by a network controller before each satellite pass. Using approximate device locations and satellite ephemeris, the controller estimates visibility windows and selects transmission parameters and start times. These assignments are delivered through an earlier downlink or configured in advance. Devices require a sufficiently accurate common time reference to follow the schedule, without continuous synchronization or a downlink immediately before each pass. We assume that this coordination is available and do not account for its signaling overhead.
\section{Simulation Results}
\label{sim}
In this section, as the first set of simulations, we evaluate the PLR of the presented system model via Monte Carlo simulations against our analytical expressions. For the simulation purposes, we set $y_c(0)=2r$. After that, we provide the PLR improvement results achieved by exploiting the RL framework in Section~\ref{RL}. Moreover, the simulation parameters are presented in Table \ref{simpar_RL}.
\begin{table}[!t]
	\renewcommand{\arraystretch}{1.15}
	\caption{Simulation and RL Parameters}
	\label{simpar_RL}
	\centering
	\footnotesize
	\begin{tabular}{@{}ll@{}}
		\toprule
		\textbf{Parameter} & \textbf{Value} \\
		\midrule
		\multicolumn{2}{@{}l}{\textit{System and channel model}}\\
		Satellite altitude ($H$) \cite{Asad_2024}                 & $500$~km \\
		Earth radius ($R$) \cite{Asad_2024}                        & $6371$~km \\
		Minimum elevation angle ($E_{\min}$)      & $1^{\circ}$ \\
		Satellite ground speed ($v$)              & $7.61$~km/s \\
		Carrier frequency, $f_c$                 & $868$ (EU), $915$ (US) MHz \\
		Number of channels                       & $64$ \\
		Payload size                             & $55$~bytes \\
		Duty cycle                               & $1\%$ \\
		Spreading factor set                     & $\{7, 8, 9, 10\}$ \\
		Coding rate set                          & $\{4/5, 4/6, 4/7, 4/8\}$ \\
		Bandwidth set                 			& $\{125, 250, 500\}$~kHz \\
		No.\ of timing options                   & $5$ \\
		\midrule
		\multicolumn{2}{@{}l}{\textit{Common learning parameters}}\\
		Network architecture                     & MLP, $1$ hidden layer ($64$, tanh) \\
		State / action dimension                 & $5$ / $240$ \\
		Discount factor               & $0.95$ \\
		Optimizer                                & Adam \\
		Training iterations                      & $60$ \\
		Held-out evaluation networks             & $5$ \\
		Random seed                              & $42$ \\
		\midrule
		\multicolumn{2}{@{}l}{\textit{PPO}}\\
		Learning rate                            & $3\times10^{-3}$ \\
		Clipping parameter           & $0.2$ \\
		GAE parameter                 & $0.95$ \\
		Epochs per update                        & $4$ \\
		Entropy coefficient                      & $0.01$ \\
		Value-loss coefficient                   & $0.5$ \\
		\midrule
		\multicolumn{2}{@{}l}{\textit{DQN}}\\
		Learning rate                            & $1\times10^{-3}$ \\
		Replay buffer size                       & $2\times10^{4}$ \\
		Mini-batch size                          & $128$ \\
		Exploration rate         & $1.0 \rightarrow 0.05$ \\
		Exploration rate-decay (iterations)         & $40$ \\
		Target-network sync (steps)              & $200$ \\
		Gradient steps per iteration             & $200$ \\
		\bottomrule
	\end{tabular}
\end{table} 

\subsection{Analytical Derivation Evaluation}
First, to evaluate the analytical expressions provided in Section~\ref{plr}, we must note that for numerical integration purposes, the upper limit of the outer integral in (\ref{PLR_Dopp_shift}), (\ref{PLR_Dopp_rate}), (\ref{q_tau}), and (\ref{PLR_col}) has been changed to $2r/v-\epsilon$ to avoid singularity. The value of $\epsilon$ has been selected within the range of $10^{-11}$ to $10^{-3}$ depending on the $\rm BW$ and $N$ via trial and error.

\begin{figure}[t!]
	\centering
	\includegraphics[width=0.6\linewidth]{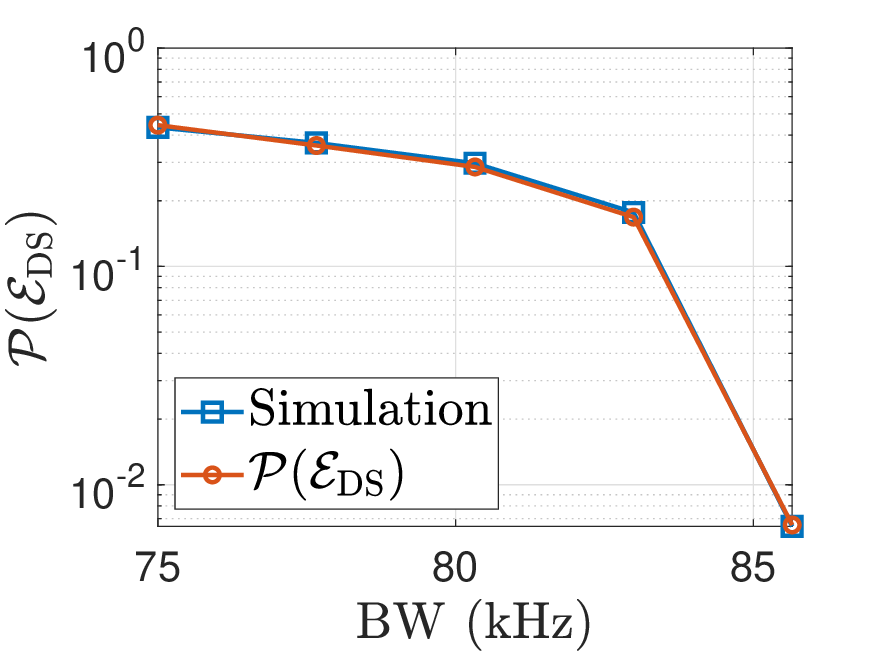}
	\caption{PLR due to Doppler shift versus $\rm BW$.}
	\label{PEDS}
\end{figure}

In Fig. \ref{PEDS}, we evaluate the PLR due to Doppler shift expression in (\ref{PLR_Dopp_shift}). It can be seen that the analytical results perfectly match the simulation.  It can also be observed that the PLR due to Doppler shift decreases as the $\rm BW$ increases. However, it can be inferred that for the operational $\rm BW$ of LoRa communication in practical settings, i.e., ${\rm BW}=125$ kHz to ${\rm BW}=500$ kHz, the PLR due to Doppler shift is almost negligible and does not have any impact on the total PLR.  

\begin{figure}[t!]
	\centering
	\includegraphics[width=0.6\linewidth]{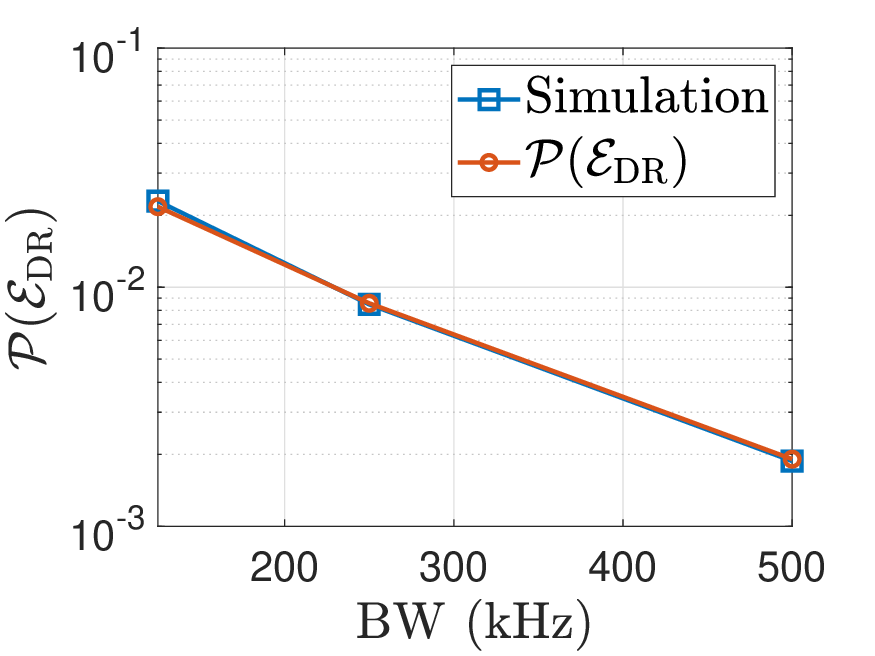}
	\caption{PLR due to Doppler rate versus $\rm BW$.}
	\label{PEDR}
\end{figure}

Fig. \ref{PEDR} shows the PLR due to Doppler rate for the practical values of $\rm BW$, i.e., $125$, $250$, and $500$ kHz. Fig. \ref{PEDR} verifies our analytical expression presented in (\ref{PLR_Dopp_rate}) and also shows that despite the Doppler shift effect, which is negligible in practical $\rm BW$ values, Doppler rate has a significant effect on PLR of the proposed network. Also, it can be seen that network performance degradation due to Doppler rate is higher for lower values of $\rm BW$. 

\begin{figure}[t!]
	\centering
	\includegraphics[width=0.6\linewidth]{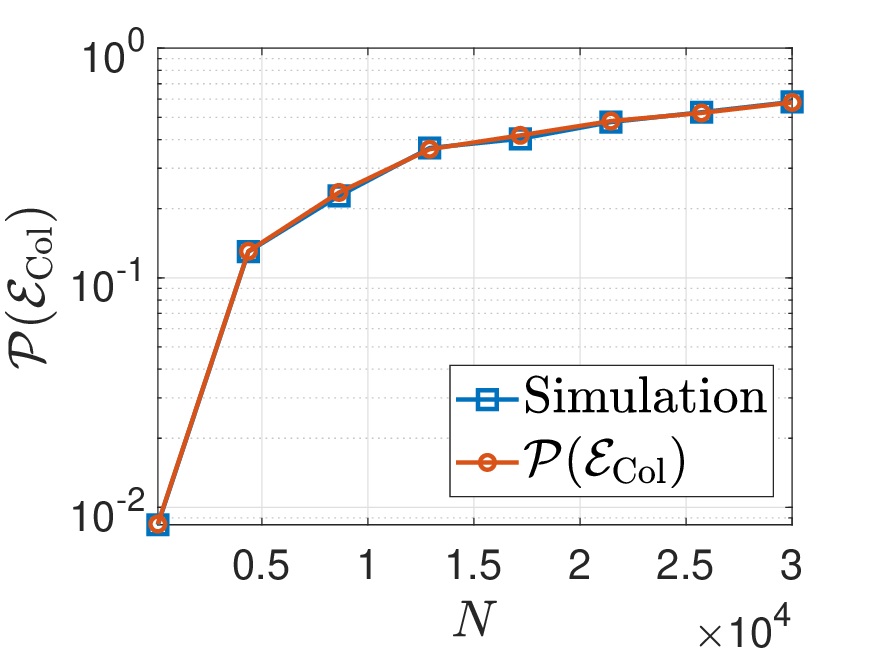}
	\caption{PLR due to collision versus $N$.}
	\label{PCol}
\end{figure}

In Fig. \ref{PCol}, PLR due to collision in the time and frequency domains is illustrated. While validating the analytical derivation in (\ref{PLR_col}), it can be seen that the random allocation of the resources such as $\rm CH$, $\rm SF$, and $\rm CR$ will result in network congestion for more than almost $4,500$ EDs (considering that in most practical cases, an acceptable PLR performance threshold is $10^{-1}$).  

\begin{figure}[t!]
	\centering
	\includegraphics[width=0.6\linewidth]{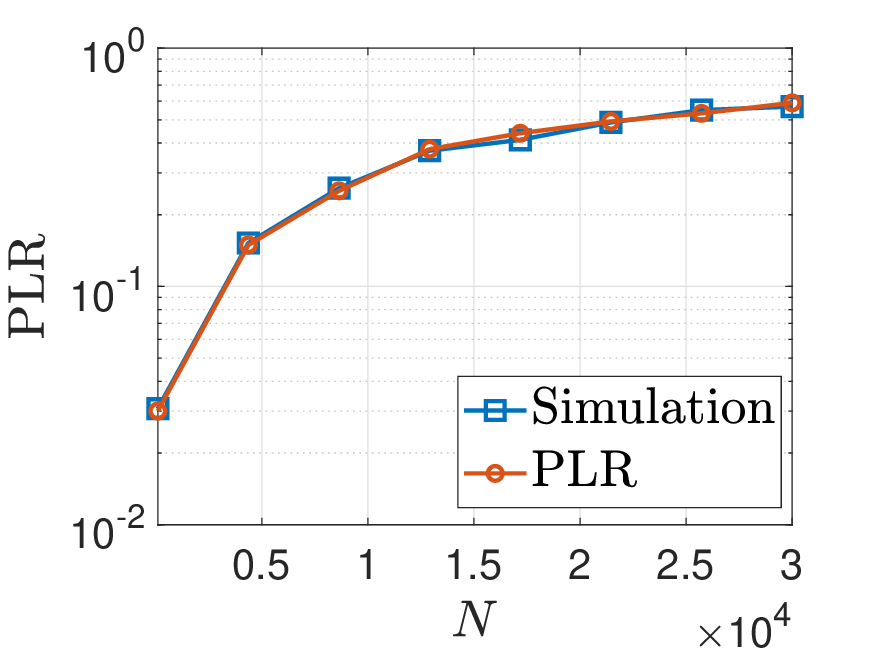}
	\caption{PLR due to collision, Doppler shift, and Doppler rate versus $N$.}
	\label{PLR_tot_fig}
\end{figure}

Fig. \ref{PLR_tot_fig} shows the combined effect of Doppler shift, Doppler rate, and collision on the network performance in terms of total PLR. It can be observed that the performance degradation follows a trend similar to that of the PLR due to collisions; however, the presence of Doppler rate effects further exacerbates the degradation. 

Having verified the analytical derivations, we now present the simulation results of the proposed RL framework, in which the analytically derived expressions are incorporated into the reward function.

\subsection{RL-based PLR Improvement Results}
We use PPO and DQN as the RL techniques to solve the optimization problem $\mathcal{P}_{\rm plr}$ in~(\ref{Opt}), as stated in Section~\ref{RL}. The agent chooses an SF, CR, and BW combination along with a packet-burst timing for each ED in both schemes after observing a per-ED state (\ref{state}). The channel is then randomly assigned among the $64$ available channels. 

It should be noted that, as defined in (\ref{action}), each ${\rm ED}_i$ has $48$ possible parameter combinations for selecting $c_i$. However, throughout the analytical derivation of the collision probability, ${\mathcal{P}(\mathcal{E}{\rm col})}$, we saw that longer packets incur a higher probability of collision. Consequently, if all $48$ combinations were always available, the RL agent would naturally converge to selecting the combination with the smallest ${\rm ToA}$ for every ED in order to maximize the reward by minimizing the PLR. Such behavior is unrealistic in practice because the transmission parameters, and therefore the resulting ${\rm ToA}$, cannot be chosen arbitrarily. Instead, they are constrained by the received signal-to-noise ratio (SNR), which is primarily determined by the distance between the ED and the GW. To reflect this practical limitation, the range between the minimum and maximum ED to GW distances is partitioned into $48$ intervals. Each interval is mapped to a minimum allowable ${\rm ToA}$, and an ED is permitted to select only from the parameter combinations whose resulting ${\rm ToA}$ is equal to or greater than this threshold. Consequently, EDs farther from the GW are restricted to longer ${\rm ToA}$ values, whereas EDs closer to the GW retain access to a larger set of shorter-${\rm ToA}$ configurations, consistent with practical LoRaWAN operation.

Through the training process, each policy is trained on independently generated network realizations and then evaluated on a set of previously unseen realizations; unless stated otherwise, all reported PLR and throughput values are averaged over $5$ held-out networks. The resulting allocations are benchmarked against the random allocation scheme of Fig. \ref{PLR_tot_fig}. 

\begin{figure}[t!]
	\centering
	\includegraphics[width=\linewidth]{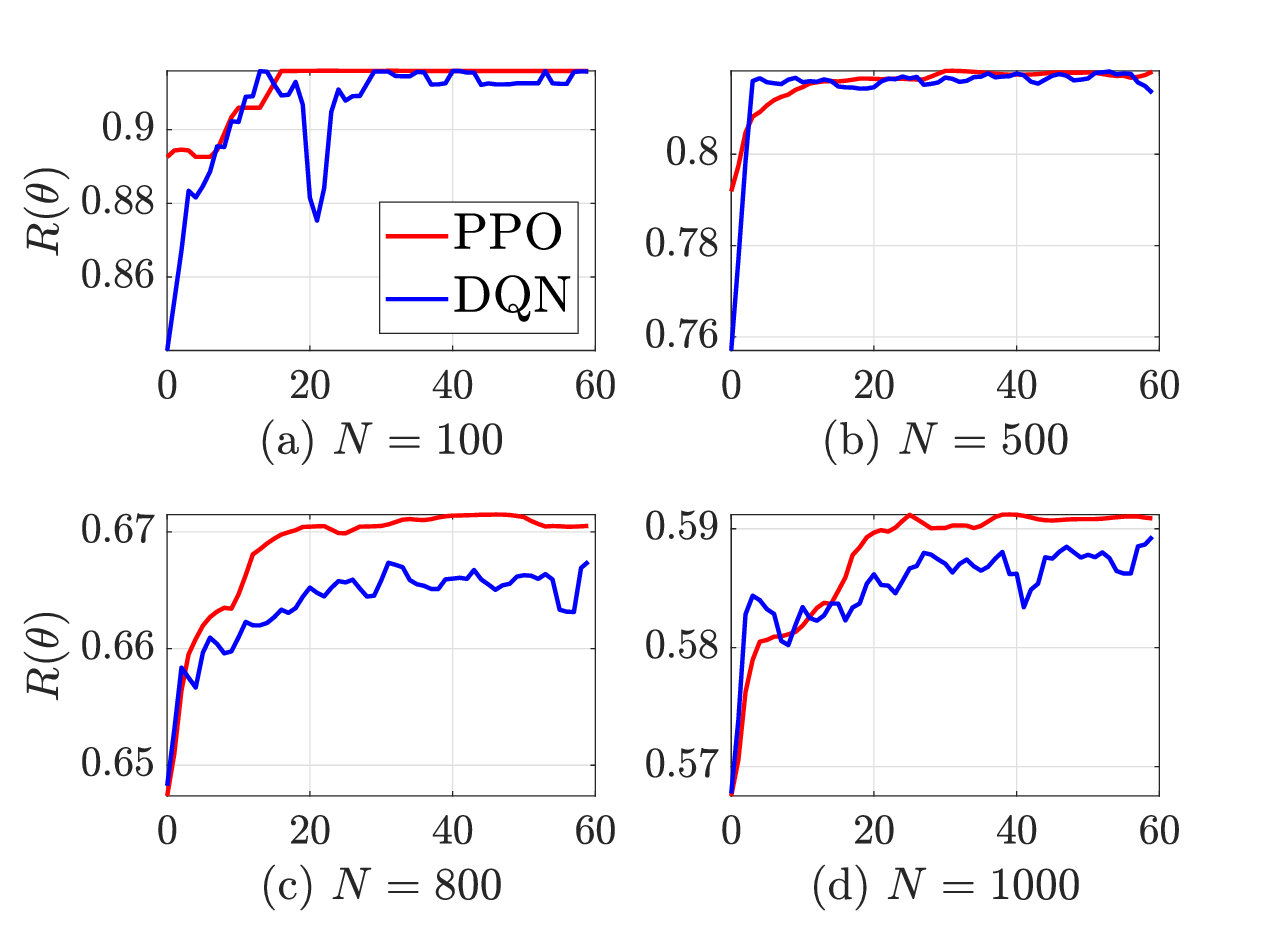}
	\caption{Training reward (mean per-ED success probability) versus training
		iteration for PPO and DQN at $N=100,500,800,1000$. Both agents converge completely within
		$\sim\!30$--$40$ iterations; DQN starts lower and catches up to PPO.}
	\label{Train}
\end{figure}

Fig.~\ref{Train} shows the training reward, defined as the mean per-ED packet success probability in (\ref{objective}), versus the training iteration for different network sizes. Both agents converge rapidly within approximately $10$--$15$ iterations, with PPO exhibiting smoother convergence and slightly higher rewards at larger $N$. DQN initially achieves lower rewards due to its randomly initialized action-value estimates but improves as the value function is learned. For both algorithms, the converged reward decreases as the network becomes denser, from approximately $0.92$ at $N=100$ to $0.59$ at $N=1000$, reflecting the increasing occurrence of unavoidable packet collisions.

\begin{figure}[t!]
	\centering
	\includegraphics[width=0.8\linewidth]{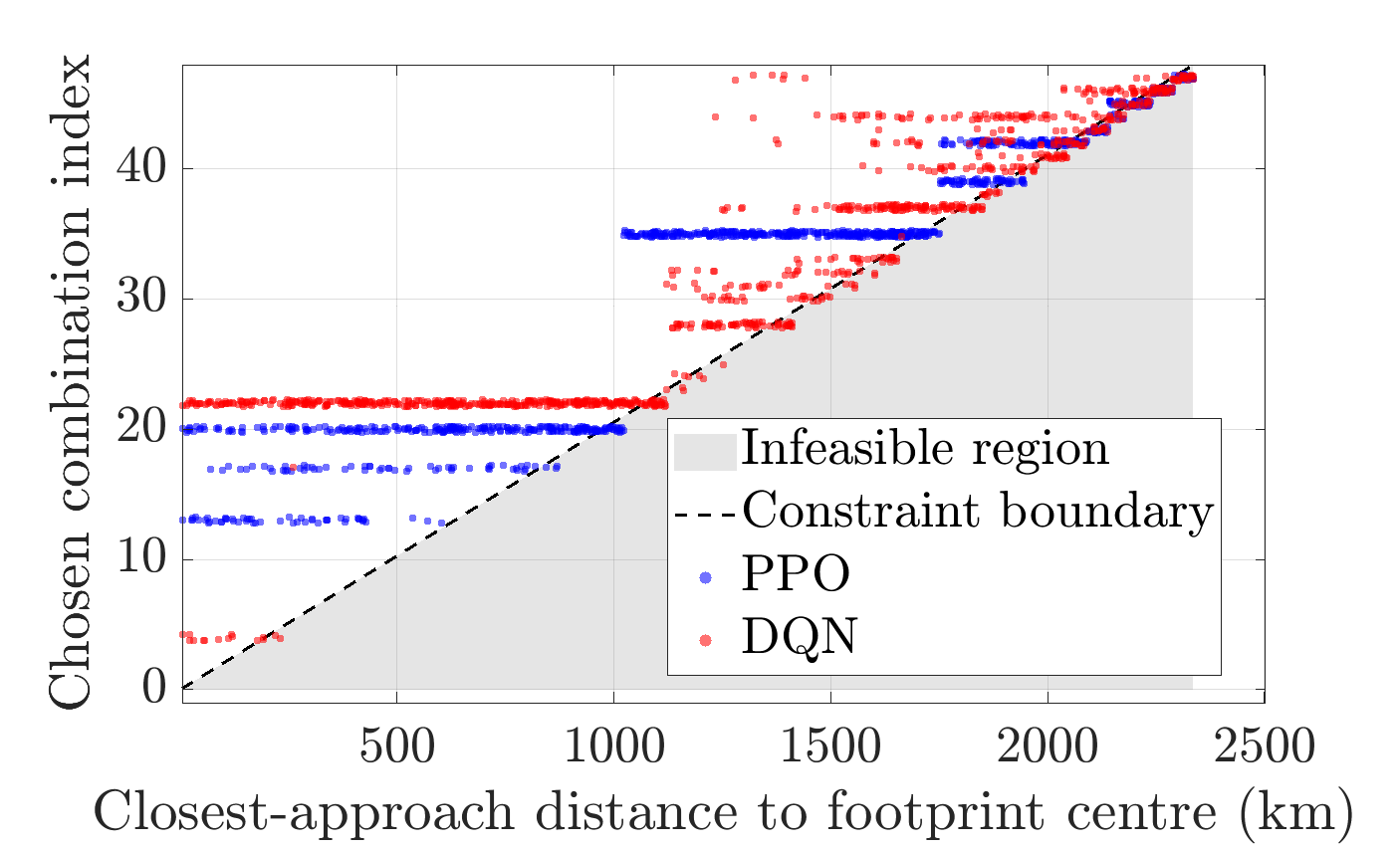}
	\caption{Chosen combination index vs.\ closest-approach distance ($N=1000$): both agents satisfy the distance constraint.}
	\label{Dist_Comb_N1000}
\end{figure}

The combination index chosen by each RL agent as a function of each ED's closest-approach distance is displayed for $N=1000$ in Fig.~\ref{Dist_Comb_N1000}. The vertical axis is the combination index $\in\{0,1,\dots,47\}$, obtained by enumerating all $4\times4\times3=48$ $(\mathrm{SF},\mathrm{CR},\mathrm{BW})$ combinations and sorting them in increasing order of ToA, ${\rm ToA}_0\le{\rm ToA}_1\le\cdots\le{\rm ToA}_{47}$. Thus, the index ranges from the shortest packet ($j=0$: ${\rm SF}=7$, ${\rm CR}=4/5$, ${\rm BW}=500$~kHz resulting in ${\rm ToA}\approx20$~ms) to the longest (${\rm SF}=10$, ${\rm CR}=4/8$,	${\rm BW}=125$~kHz resulting in ${\rm ToA}\approx721$~ms), so that a lower index corresponds to a shorter, higher-rate, wider-bandwidth packet and a higher index to a longer, more robust one. The feasibility border is defined as $\lfloor 48\,d/d_{\max}\rfloor$ where $d_{\max}=r$ is the maximum distance of an ED from the satellite footprint center during its visibility duration. This is represented by the solid line, and infeasible actions are represented by the shaded area below it. By means of Fig. \ref{Dist_Comb_N1000}, two key observations can be made. First, the action-masking mechanism effectively stops the agents from assigning transmission configurations with a ${\rm ToA}$ shorter than that permitted by the communication distance, as it is clear that all selections made by both PPO and DQN lie on or above this boundary, with no points appearing in the infeasible region. Second, it can be observed that many of the chosen actions for distant EDs lie directly on the feasibility boundary because the received SNR decreases with increasing closest-approach distance, which reduces the number of feasible transmission combinations and forces EDs to choose configurations with progressively longer ${\rm ToA}$. Both agents primarily choose low-index combinations within the feasible region, which correspond to shorter ${\rm ToA}$ and larger bandwidths. Such configurations reduce the channel occupancy time and, consequently, the likelihood of packet collisions, while the use of larger bandwidths also improves robustness against Doppler-induced impairments, based on the results of Fig. \ref{PEDS} and Fig. \ref{PEDR}.

\begin{figure}[t!]
	\centering
	\includegraphics[width=1.05\linewidth]{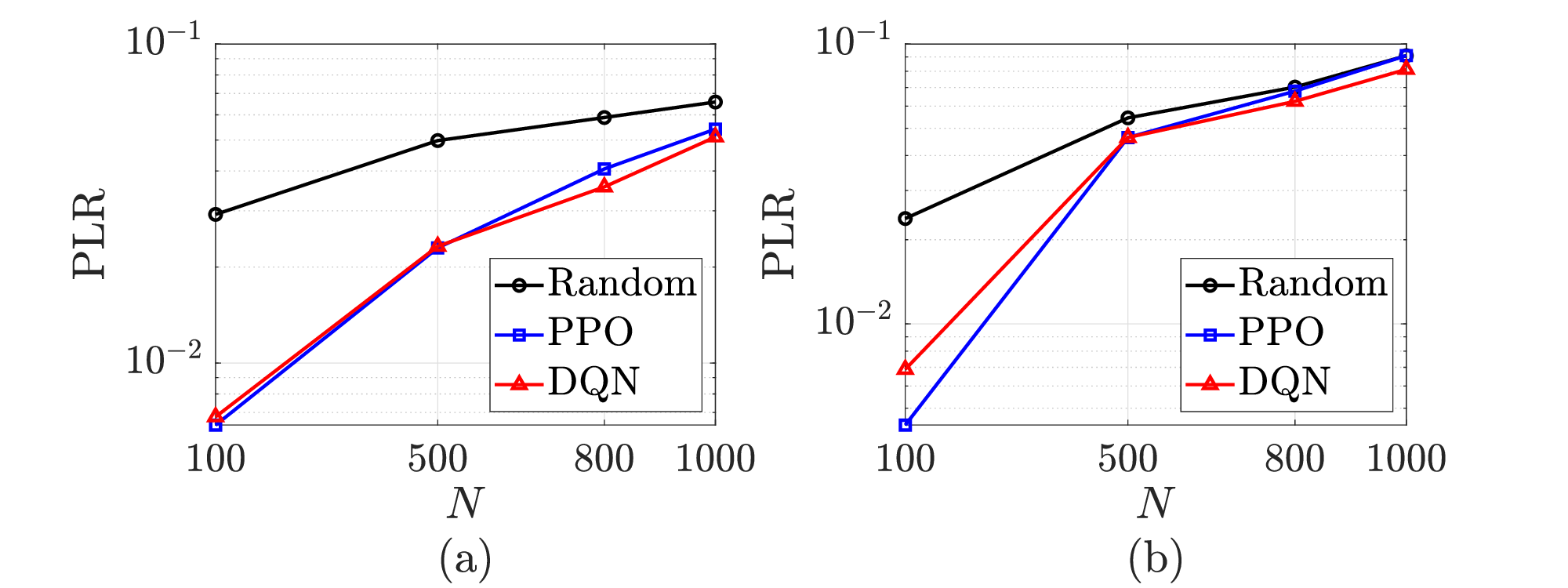}
	\caption{PLR performance of PPO and DQN approaches compared to random resource allocation scheme for (a) a duty cycle of $1\%$ in EU region and (b) a dwell time of $400$ ms in US region.}
	\label{PLR_RL}
\end{figure}

The total PLR of the two learned policies and the random allocation is shown in Fig.~\ref{PLR_RL} (a) as the number of EDs increases from $100$ to $1000$. The PLR is significantly reduced by both PPO and DQN at all network sizes, and the gain is most naturally described as an increase in the number of devices that can be served by a single GW installed on a LEO satellite. It can be seen that for a target PLR of $3\times10^{-2}$, PPO and DQN approximately enable serving $600$ and $650$ EDs, respectively, which is around $600\%$ and $650\%$ more network capacity than that of random resource allocation. It should be noted that the $1\%$ duty-cycle assumption considered so far is commonly associated with European sub-bands. We additionally evaluate a U.S. region scenario with a maximum channel occupancy of $400$~ms within $20$~s \cite{FCC_reg}. For the considered $55$-byte payload, this constraint excludes ${\rm SF}=10$ configurations. Fig.~\ref{PLR_RL} (b) compares PPO, DQN, and random resource allocation under this restriction. Both learned policies substantially reduce the PLR at $N=100$, but their advantage decreases as the network becomes denser. At larger network sizes, PPO approaches random allocation, while DQN retains a modest performance gain. 

%
	
\begin{figure}[t!]
	\centering
	\includegraphics[width=0.6\linewidth]{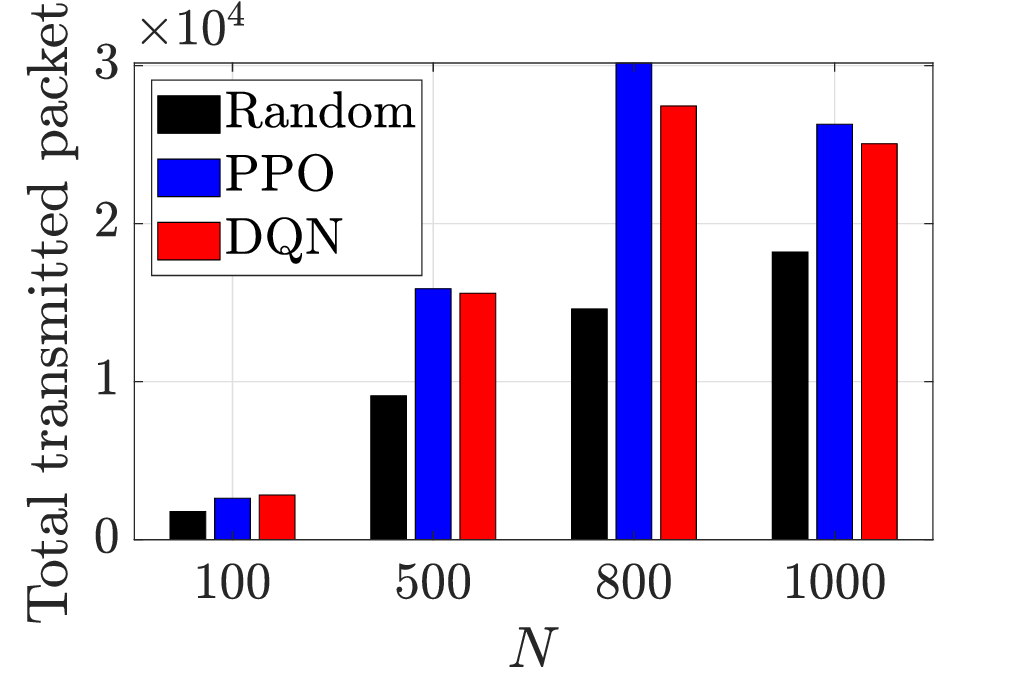}
	\caption{Total number of transmitted packets versus the number of EDs $N$ for the
		random allocation, PPO, and DQN.}
	\label{Num_Pack_RL}
\end{figure}

Finally, in Fig. \ref{Num_Pack_RL}, we can see that both RL schemes resulted in more total transmitted packets by all EDs compared to the random resource allocation. The reason behind this is that RL agents learned that lower $\rm ToA$ values resulted in lower collision probability. This inherently means more available transmission time for each ED, boosting the network's overall number of packets transmitted. The summarized network total numbers of transmitted packets are given in Table \ref{tab_Num_pack_RL}
\begin{table}[!t]
	\renewcommand{\arraystretch}{1.2}
	\caption{Total Transmitted Packets and Throughput Gain Over the Random
		Allocation}
	\label{tab_Num_pack_RL}
	\centering
	\begin{tabular}{@{}ccccccc@{}}
		\toprule
		\multirow{2}{*}{$N$} & \multicolumn{3}{c}{Total transmitted packets} &
		& \multicolumn{2}{c}{Gain over random} \\
		\cmidrule(lr){2-4}\cmidrule(lr){6-7}
		& Random & PPO & DQN & & PPO & DQN \\
		\midrule
		$100$  & $1{,}778$  & $2{,}612$  & $2{,}819$  & & $+46.9\%$  & $+58.5\%$ \\
		$500$  & $9{,}096$  & $15{,}877$ & $15{,}595$ & & $+74.5\%$  & $+71.4\%$ \\
		$800$  & $14{,}595$ & $30{,}164$ & $27{,}445$ & & $+106.7\%$ & $+88.0\%$ \\
		$1000$ & $18{,}200$ & $26{,}285$ & $25{,}054$ & & $+44.4\%$  & $+37.7\%$ \\
		\bottomrule
	\end{tabular}
\end{table}	
	
\section{Conclusion}
\label{conc}
In this paper, we investigated resource allocation in DtS-IoT LoRaWAN by jointly considering time- and frequency-domain packet collisions and Doppler impairments caused by satellite motion. We first derived an analytical PLR model, which was validated through Monte Carlo simulations and subsequently employed to construct the reward function for PPO- and DQN-based resource allocation. Simulation results demonstrated that both RL schemes significantly outperform random allocation. At a target PLR of $3\times10^{-2}$, PPO and DQN support approximately $650$ and $690$ EDs, respectively, compared with only $100$ EDs under random allocation. Moreover, for $800$ EDs, PPO and DQN increase the total number of transmitted packets during a single satellite pass by $106.7\%$ and $88.0\%$, respectively. These results demonstrate the potential of combining analytical network modeling with RL-based resource allocation to improve the scalability and reliability of DtS-IoT LoRaWAN. Future work will consider multi-satellite scenarios, dynamic traffic conditions, and multi-agent RL for distributed resource allocation in large-scale satellite IoT networks.

\appendices
\section{Proof of lemma \ref{lemmCDF}}
\label{appA}
First, to derive $F_{Y|X}(y|x)$, we use the following relation between the PDFs:
\begin{IEEEeqnarray}{rCl}
\label{f_YX}
f_{Y|X}(y|x) = \frac{f_{X,Y}(x,y)}{f_X(x)}.
\end{IEEEeqnarray}
 Considering the uniform distribution of $x$ and $y$, it is straightforward to show that 
\begin{IEEEeqnarray}{rCl}
	\label{fxy}
	f_{X,Y}(x,y)=\frac{1}{4ry_c(0)+\pi r^2}.
\end{IEEEeqnarray}
Moreover, since $-r\leq x \leq r$, we can write the minimum and maximum values that $y$ can take in terms of $x$ as $y_{\rm min} = -y_c(0)-\sqrt{r^2-x^2}$ and $y_{\rm max} = y_c(0)+\sqrt{r^2-x^2}$. Based on this, the PDF of $x$ can be simply written as the length of segment bounded by $y_{\rm min}$ and $y_{\rm max}$ divided by the total area of the coverage region, i.e.,
 \begin{IEEEeqnarray}{rCl}
 	\label{fx_app}
 	f_X(x) = \frac{y_{\rm max}-y_{\rm min}}{4ry_c(0)+\pi r^2}.
 \end{IEEEeqnarray}
Therefore, by substituting (\ref{fxy}) and (\ref{fx_app}) into (\ref{f_YX}) and performing an integration, we obtain:
\begin{IEEEeqnarray}{rCl}
	\label{CDF_y_app}
	F_{Y|X}(y|x) = \frac{y-y_{\rm min}}{2y_c(0)+2\sqrt{r^2-x^2}},
\end{IEEEeqnarray}

Next, the CDF of $\xi_n(t)$ can be expressed as:
\begin{IEEEeqnarray}{rCl}
	\label{Prob_CDF_Dopp}
	{\rm Pr}\left\{\xi_n(t)\leq f_{\rm th}\right\}.
\end{IEEEeqnarray}
 On the other hand, we can rewrite $\gamma_n(t)$ in terms of $x$-coordinate and $y$-coordinate as:
 \begin{IEEEeqnarray}{rCl}
 	\label{gamma_xy}
 	\gamma_n(t)=2\sin^{-1}\left[\frac{x^2+(y-y_c(0)+vt)^2}{2R}\right],
 \end{IEEEeqnarray}
 and by taking the derivative with respect to $t$, we have
 \begin{IEEEeqnarray}{rCl}
 	\label{gamma_min_x}
 	\gamma_{{\rm min},n}= 2\sin^{-1}\left(\frac{|x|}{2R}\right).
 \end{IEEEeqnarray}
 
 Hence, (\ref{Prob_CDF_Dopp}) can be rewritten as (\ref{Prob_CDF_Dopp_xy}) at the top of the next page.
 \begin{figure*}
 	\normalsize
 \begin{IEEEeqnarray}{rCl}
 	\label{Prob_CDF_Dopp_xy}
 	{\rm Pr}\left\{\xi_n(t)\leq f_{\rm th}\right\}={\rm Pr}\left\{\sqrt{\frac{{\cos^2\left[2\sin^{-1}\left(\frac{|x|}{2R}\right)\right]-\cos^2\left[2\sin^{-1}\left[\frac{x^2+(y-y_c(0)+vt)^2}{2R}\right]\right]}}{{1+K^2-2K\cos\left[2\sin^{-1}\left[\frac{x^2+(y-y_c(0)+vt)^2}{2R}\right]\right]}}\leq \frac{f_{\rm th}}{\rho}}\right\}.
 \end{IEEEeqnarray}
 \hrulefill
 \vspace
 *
 {4pt}
 \end{figure*}
  After mathematical simplifications and using $\cos[2\sin^{-1}(x)]=1-2x^2$, we can write
\begin{IEEEeqnarray}{rCl}
 		\label{Prob_CDF_Dopp_xy_simp}
 		{\rm Pr}\left\{\xi_n(t)\leq f_{\rm th}\right\}={\rm Pr}\left\{\frac{a_1 y'^4+a_2 y'^2}{a_3+a_4 y'^2}\leq a_5^2\right\},
 	\end{IEEEeqnarray}
 where $y'=y-y_c(0)+vt$. By defining $y''=y'^2$, the probability in (\ref{Prob_CDF_Dopp_xy_simp}) can be expressed as:
 \begin{IEEEeqnarray}{rCl}
 	\label{Prob_CDF_Dopp_xy_simp2}
 	{\rm Pr}\left\{\xi_n(t)\leq f_{\rm th}\right\}={\rm Pr}\left\{y''\geq R_1 \;\&\; y''\leq R_2\right\}.
 \end{IEEEeqnarray}
To obtain the final analytical expression for (\ref{Prob_CDF_Dopp_xy_simp2}), we apply marginalization over the $x$-coordinate as it is a function of both $x$ and $y$. Specifically, the conditional probability is first evaluated for a given $x$ and then averaged with respect to the marginal distribution of $X$. This yields (\ref{CDF_delt}) and completes the proof of Lemma \ref{lemmCDF}.

\section{Proof of lemma \ref{lemmPDF}}
 First, we write the expression for $t_{\rm s}$ and $t_{\rm v}$ according to (\ref{dn}), (\ref{tsn}), and (\ref{Tvn}) as:
 \begin{IEEEeqnarray}{rCl}
 	\label{ts_app}
 	t_{{\rm s},n}=\begin{cases}
 		0, & \text{if $(x,y)\in$ Area (I)},\\
 		\frac{y_c(0)-y-\sqrt{r^2-x^2}}{v}, & \text{o.w.},
 	\end{cases}.
 \end{IEEEeqnarray}
 and
 \begin{IEEEeqnarray}{rCl}
 	\label{tv_app}
 	t_{{\rm v},n}=\begin{cases}
 		\frac{y_c(0)-y+\sqrt{r^2-x^2}}{v}, & \text{if $(x,y)\in$ Area (I)},\\
 		\frac{2\sqrt{r^2-x^2}}{v}, & \text{if $(x,y)\in$ Area (II)},\\
 		\frac{y_c(0)+y+\sqrt{r^2-x^2}}{v}, & \text{if $(x,y)\in$ Area (III)}.
 	\end{cases}
 \end{IEEEeqnarray}

It can be seen that within Area (II), we have $t_{{\rm s},n}=[y_c(0)-y]/v-t_{{\rm v},n}/2$. Also, in Area (III), we have $t_{{\rm s},n}+t_{{\rm v},n}=2y_c(0)/v$. As a result, the joint distribution will not be completely continuous and it has a singular component from Area (I) along $t_{{\rm s},n}=0$, a 2D continuous density from Area (II), and another singular component from Area (III) along $t_{{\rm s},n}=2y_c(0)/v-t_{{\rm v},n}$. We can write:
 \begin{IEEEeqnarray}{rCl}
 	\label{ftstv}
 	f_{T_{\rm v},T_{\rm s}}(t_{{\rm v},n},t_{{\rm s},n})&=&f_1(t_{{\rm v},n},t_{{\rm s},n})+f_2(t_{{\rm v},n},t_{{\rm s},n})\nonumber\\
 	&&+f_3(t_{{\rm v},n},t_{{\rm s},n}),
 \end{IEEEeqnarray}
which breaks down the total joint PDF for three different areas. 

First, we derive $f_1(t_{{\rm v},n},t_{{\rm s},n})$. In this case, we know that $t_{{\rm s},n}$ is deterministic and equals to zero. So, we compute the marginal density of $t_{{\rm v},n}$ as
\begin{IEEEeqnarray}{rCl}
	\label{f1_tv}
	f_1(t_{{\rm v},n},t_{{\rm s},n})=\int_{-\sqrt{r^2-\frac{v^2 t_{{\rm v},n}^2}{4}}}^{\sqrt{r^2-\frac{v^2 t_{{\rm v},n}^2}{4}}} f_{X,Y}(x,y)\Big{|}\frac{\partial y}{\partial t_{\rm v}}\Big{|}{\rm d}x,
\end{IEEEeqnarray}
where inside Area (I), $y=y_c(0)+\sqrt{r^2-x^2}-vt_{{\rm v},n}$. This yields:
\begin{IEEEeqnarray}{rCl}
	\label{f1_tv_fin}
	f_1(t_{{\rm v},n},t_{{\rm s},n})=\frac{2v\sqrt{r^2-\frac{v^2t_{{\rm v},n}^2}{4}}}{4ry_c(0)+\pi r^2}\delta(t_{{\rm s},n}).
\end{IEEEeqnarray}

Next, we calculate $f_2(t_{{\rm v},n},t_{{\rm s},n})$. The $x$ and $y$ in Area (II) can be written as $x=\pm\sqrt{r^2-v^2t^2_{{\rm v},n}/4}$ and $y=y_c(0)-vt_{{\rm v},n}-vt_{{\rm s},n}/2$ with supports $0\leq t_{{\rm v},n}\leq 2r/v$ and $0\leq t_{{\rm s},n}\leq 2y_c(0)/v-t_{{\rm v},n}$. Also, we can simply write the following using Jacobian theorem for PDFs:
\begin{IEEEeqnarray}{rCl}
	\label{one_to_many}
	f_2(t_{{\rm v},n},t_{{\rm s},n}) = \sum_{i=1}^2 f_{X,Y}(x,y)|J_i|,
\end{IEEEeqnarray}
where the summation appears due to the fact that both $+x$ and $-x$ map to the same point $(t_{{\rm v},n},t_{{\rm s},n})$. Also, the Jacobian magnitudes for both branches are equal and can be calculated as 
\begin{IEEEeqnarray}{rCl}
	\label{jacob}
	|J_1|=|J_2|=\frac{v^3t_{{\rm v},n}}{4\sqrt{r^2-\frac{v^2t^2_{{\rm v},n}}{4}}}.
\end{IEEEeqnarray}
Hence, we have:
\begin{IEEEeqnarray}{rCl}
	\label{f2_tv_fin}
	f_2(t_{{\rm v},n},t_{{\rm s},n})=\frac{v^3t_{{\rm v},n}}{2(4ry_c(0)+\pi r^2)\sqrt{r^2-\frac{v^2t_{{\rm v},n}^2}{4}}}.
\end{IEEEeqnarray}

Finally, $f_3(t_{{\rm v},n},t_{{\rm s},n})$ can be derived using a similar approach to $f_1(t_{{\rm v},n},t_{{\rm s},n})$. Only with the difference that inside Area (III), $y=-y_c(0)-\sqrt{r^2-x^2}+vt_{{\rm v},n}$. Hence, we have 
\begin{IEEEeqnarray}{rCl}
	\label{f3_tv_fin}
	f_3(t_{{\rm v},n},t_{{\rm s},n})=\frac{2v\sqrt{r^2-\frac{v^2t_{{\rm v},n}^2}{4}}}{4ry_c(0)+\pi r^2}\delta\left(t_{{\rm s},n}+t_{{\rm v},n}-\frac{2y_c(0)}{v}\right),\nonumber\\
\end{IEEEeqnarray}
which completes the proof. 
\bibliographystyle{IEEEtran}
\bibliography{IoT_Transmission_Techs_PhD_V20}
	
\end{document}